\pdfoutput=1
\documentclass[journal]{IEEEtran}
\IEEEoverridecommandlockouts
\usepackage[cmex10]{amsmath}
\usepackage[noadjust]{cite}
\usepackage{%
    amssymb,%
    color,%
    enumerate,%
    etoolbox,%
    mathtools,%
    stmaryrd%
}
\usepackage{subcaption}
\usepackage[ruled,vlined]{algorithm2e}
\usepackage{authblk}
\usepackage{tikz,pgf,pgfplots}
\pgfplotsset{compat=newest}
\pgfplotsset{plot coordinates/math parser=false}
\usepackage{helvet}
\usepackage[eulergreek]{sansmath}
\pgfplotsset{
  tick label style = {font=\sansmath\sffamily\footnotesize},
  legend style = {font=\sansmath\sffamily\footnotesize},
  label style = {font=\sansmath\sffamily\footnotesize}
}
\newlength\figureheight
\newlength\figurewidth

\newcommand{\CE}{\color{black}} %End of Change
\begin{document}

\title{Scheduling and Power Control for Wireless Multicast Systems via Deep Reinforcement Learning\large{${}^\dagger$}\thanks{${}^\dagger$Preliminary version of a part of this paper was presented in Allerton, 2019.}}
\author[1]{Ramkumar Raghu} 
%\author[1]{Pratheek Upadhyaya} 
\author[1]{Mahadesh Panju} 
\author[1,2]{Vaneet Aggarwal} 
\author[1]{Vinod Sharma}  
\affil[1]{Indian Institute of Science, Bangalore, INDIA. \textit{\{ramkumar,mahadesh,vinod\}@iisc.ac.in}}
\affil[2]{Purdue University, West Lafayette IN, USA.  \textit{vaneet@purdue.edu}}

\maketitle

\begin{abstract} 
 
%Multicasting in wireless systems is being studied with a renewed interest as it is seen as a natural way to exploit the redundancy in user requests in a Content Centric Network. Power control and optimal scheduling of transmissions are paramount for QoS improvement in wireless multicast networks. We consider a single hop multicast scheme recently proposed for wireless downlinks in \cite{wcnc}. Several queuing strategies were demonstrated to improve performance of the multicast network under fading. It was also shown that power control can significantly improve the multicast network's performance. However there are two main drawbacks of the proposed schemes. First, the optimality of each queueing strategy depends on the system parameters such as arrival rates, popularity etc., and there is no one-scheme-fits-all solution for variations in these system parameters. Secondly the  proposed scheme for power control is not scalable and obtaining optimal power control is intractable because of very large state space. 
Multicasting in wireless systems is a natural way to exploit the redundancy in user requests in a Content Centric Network. Power control and optimal scheduling can significantly improve the wireless multicast network’s performance under fading. However, the model-based approaches for power control and scheduling studied earlier are not scalable to large state space or changing system dynamics. 
In this paper, we use deep reinforcement learning where we use function approximation of the Q-function via a deep neural network to obtain a power control policy that matches the optimal policy for a small network. We show that power control policy can be learnt for reasonably large systems via this approach. Further we use multi-timescale stochastic optimization to maintain the average power constraint. We demonstrate that a slight modification of the learning algorithm allows tracking of time varying system statistics. Finally, we extend the multi-time scale approach to simultaneously learn the optimal queueing strategy along with power control. We demonstrate scalability, tracking and cross-layer optimization capabilities of our algorithms via simulations. The proposed multi-time scale approach can be used in general large state-space dynamical systems with multiple objectives and constraints, and may be of independent interest.    
 
\end{abstract}
\section{Introduction}

Content services such as Netflix, Prime Video, etc., have dramatically increased the demand for high-definition videos over mobile networks. Almost $78\%$ of mobile data traffic is expected to be due to these mobile videos \cite{CVNI}. It is observed that the request traffic for these contents have multiple redundant requests \cite{itube1}. Next generation wireless networks are being constantly upgraded to satisfy  these exploding demands by exploiting the nature of the request traffic.  Serving the redundant requests simultaneously is a natural way to utilize network resources efficiently. Thus, efficient multicasting is studied widely in the wireless networking community.

A multicast queue with network coding is studied in \cite{Moghadam} with an infinite library of files. The case of slotted broadcast systems with one server transmitting to multiple users is studied in \cite{Cogil}. Some recent works \cite{Maddah-Ali2014} use coded caching to achieve multicast. This approach uses local information in the user caches to decode the coded transmission and provides improvement in throughput by increasing the effective number of files transferred per transmission. This throughput may get reduced in a practical scenario due to queueing delays at the basestation/server. \cite{Rezaei2016a} addresses these issues, analyses the queuing delays and compares it with an alternate coded scheme with LRU caches (CDLS) which provides improvement over the coded schemes in \cite{Maddah-Ali2014}. A more recent work in this direction, \cite{Arxiv2018} provides alternate multicast schemes and analyses queueing delays for such multicast systems. In \cite{Arxiv2018}, it is shown that a simple multicast scheme, can have significant gains over the schemes in \cite{Maddah-Ali2014}, \cite{Rezaei2016a} in high traffic regime. 

We further study the multicast scheme proposed in \cite{Arxiv2018} in this paper. This multicast queue merges the requests for a given file from different users, arriving during the waiting time of the initial requests. The merged requests are then served simultaneously.  The gains achieved by this simple multicast scheme, however, are quickly lost in wireless channels due to fading. It suffers from the users with bad channels, thereby decreasing the QoS even for users with good channels. The authors of \cite{wcnc} studied this problem and proposed novel schemes, which provide significant multicast gains under fading as compared to the simple multicast. Further, it was shown that an optimal state dependent power control can significantly improve the average delays experienced by the users. 

The queueing schemes and the power control policy proposed in \cite{wcnc}, though provide improved delays, have following limitations. 1) The queueing scheme which performs best depends on the system parameters such as size of the system, the request rate, etc. 2) The algorithm to obtain the power control policy is not scalable with the number of users and the number of states of the channel gains. Also, the policy doesn't adapt to changing system statistics, which in turn depends on the power control policy. 3) The queueing schemes and power control are dealt individually. This paper tries to overcome the above limitations of the scheme in \cite{wcnc}.%Therefore, the queueing and the power control need to be optimized together and need to track the changing system dynamics.

%\begin{itemize}
%\item The optimality of each queueing scheme depends on the system parameters such as size of the system, the request rate, etc.
%\item The power control algorithm to obtain the policy is not scalable with the number of users and the number of states of the channel gains.
%\item The policy doesn't adapt to changing system statistics, which in turn depends on the policy.
%\item The queueing schemes and power control are dealt individually. We will see that the optimality of queuing schemes and the power control policy are interdependent, hence, requires simultaneous optimization.
%\end{itemize}           

We first provide algorithms for the two optimization problems individually and then combine the two algorithms to obtain the overall optimal queuing strategy and the power control. Stochastic optimization (\cite{borkar}) is a useful tool to obtain the optimally parametrized queuing strategy. However, for convergence of stochastic optimization algorithms, a careful approximation of stochastic gradients is necessary. One challenge here is that the cost to be optimized is the mean stationary sojourn time of requests to be delivered. We propose a new Deep Assisted Gradient Approximation algorithm, where, the novelty is in deriving the gradients from a Deep Network assisted by a memory. This memory helps retain the history of the explored regions and also allows adaptation to changing system dynamics in an online fashion. The replay memory and online training of the deep network adds an important feature called Importance Sampling to the stochastic optimization, which improves the confidence (lower variance) in the gradient descent steps\CE.      

Multicast systems with power control can be conveniently modeled as Markov Decision Process (MDP) but with large state and action spaces. Obtaining transition probabilities and the optimal policy, however, for such large MDPs is not feasible. Reinforcement learning, particularly, Deep reinforcement learning \cite{DBLP:journals/corr/abs-1810-06339}, comes as a natural tool to address such problems. Reinforcement learning can be used even when the transition probabilities are not available. However, large state/action space can still be an issue. Using function approximation via deep neural networks can provide significant gains. Several deep reinforcement learning techniques such as Deep Q-Network \cite{Mnih2015}, Trust Region Policy Optimization (TRPO) \cite{DBLP:journals/corr/SchulmanLMJA15}, Proximal Policy Gradient (PPO) \cite{DBLP:journals/corr/SchulmanWDRK17}, etc. have been successfully applied to several large state-space dynamical systems such as Atari \cite{DBLP:journals/corr/MnihKSGAWR13}, AlphaGo \cite{Silver1140}, etc. DQN is based on value iteration. TRPO and PPO are policy gradient based methods. Policy-Gradient methods often suffer from high variance in sample estimates and poor sample efficiency \cite{DBLP:journals/corr/abs-1810-06339}. Value iteration based deep RL methods, like DQN, have been theoretically shown to have better performance \cite{DBLP:journals/corr/abs-1901-00137} due to target network and Replay memory and provide global minimum.  	
%\indent In addition to the above mentioned trade-offs, a constrained stochastic optimization problem, as considered in this paper, further adds to the complexity of the problem. A modification of TRPO for constrained optimization is Constrained Policy Optimization \cite{DBLP:journals/corr/AchiamHTA17}. But, this too suffers from the high estimator variance issue. Work in \cite{RCPO} considers a multi-timescale approach for constrained DeepRL problems, as considered in this paper. However, \cite{RCPO} does not track the system statistics and hence cannot be applied in practical systems. 

We propose a constrained optimization variant of DQN based on multi-timescale stochastic gradient descent \cite{borkar} for power control which can track the system statistics. Finally, we develop an algorithm which combines the above two algorithms to obtain an optimal queuing strategy and power control policy. %We have preferred DQN in this work, as the Target network and Replay memory used in the DQN reduce the estimator variance and finally achieve the global minimum empirical risk \cite{DBLP:journals/corr/abs-1901-00137}. 
	 
	The major contributions of this paper are as follows:
\begin{itemize}
	\item A novel deep assisted stochastic gradient descent (DSGD) algorithm for obtaining the best queueing strategy from a given set. 
	\item Proposing two modifications to DQN to accommodate constraints and system adaptations. The constraints can be met by using a Lagrange multiplier. The appropriate Lagrange multiplier is also learnt via a two time scale stochastic gradient descent. We call this algorithm Adaptive Constrained DQN (AC-DQN). 
	\item Unlike DQN, AC-DQN can be applied to the multicast systems with constraints, as in \cite{wcnc}, to learn the power control policy, online. The proposed method meets the average power constraint while achieving the global optima as achieved by the static policy proposed in \cite{wcnc} for a small scale setup of the problem. 
	\item We demonstrate the scalability of our algorithms with system size (number of users, arrival rate, complex fading).
	\item We show that AC-DQN can track the changes in the dynamics of the system, e.g., change of rate of arrival over the time of a day, and achieve optimal performance.
	\item Finally, using the above two algorithms, we propose a generalized algorithm called Integrated DSGD and AC-DQN (IDA) to optimize systems with multiple objectives and constraints. Particularly, this algorithm is useful in any wireless network with cross-layer objectives, such as ours. IDA is a three time scale stochastic optimization algorithm for obtaining both the queuing strategy (unconstrained network layer objective) and power control (constrained physical layer objective), simultaneously. %This demonstrates how a large state-space dynamical system with multiple objectives and constraints, in general, can be optimized using multi-timescale Deep learning algorithms. 
	%\item We show that  DSGD and AC-DQN can be combined into a three time scale stochastic optimization algorithm for obtaining both the queuing strategy and power control, simultaneously. By this, we demonstrate how a large state-space dynamical system with multiple objectives and constraints, in general, can be optimized using multi-timescale Deep learning algorithms.
\end{itemize}

%\subsection{Related Work}
%\label{sec:related_work}

%Model based algorithms (where transition probabilities of the MDP are known) to obtain the optimal policy for Constrained MDPs are well established \cite{Altman}. However these algorithms are not scalable to systems with large state-space. For such systems, several model-free deep (reinforcement learning) algorithms for constrained MDPs are proposed in the literature \cite{DBLP:journals/corr/AchiamHTA17, RCPO, SafeCPO}. In the next section we point to some limitations of these algorithms such as tracking ability, computational complexity, estimator variance etc., when applied to practical systems. Our algorithms, particularly IDA, is capable of optimizing any multi-objective system, while addressing the above mentioned issues. 

We show via simulations that our algorithms choose the optimal policy among the given set of policies. Also, the power control policy obtained via our algorithm improves the delay performance of the multicast network by more than $50\%$ compared to the constant power policy. Our algorithms work equally well when we replace DQN with its improvements such as DDQN \cite{ddqn}. In fact we have run our simulations with DDQN variant of AC-DQN and have achieved similar performance.  

\subsection{Related Works}

\CE\textbf{Queueing and  Power control in Multicast Systems:} Multicast Queue and Scheduling has been studied in \cite{Mog2018,Mog2018_2,Moghadam,DelEnergyTradeOffQ}. The works in \cite{Mog2018,Mog2018_2,Moghadam} propose schemes for network coded multicast systems and analyse stability of the proposed multicast queues. Unlike these works, we use, as in our previous work \cite{Arxiv2018,wcnc}, a simple uncoded multicast queue which is always stable. In \cite{Arxiv2018}, we show that our queueing schemes perform much better than the coded multicast schemes in high traffic regimes. In the current work, we improve over the results in \cite{Arxiv2018} and \cite{wcnc} by providing novel deep learning based queueing strategies. \cite{DelEnergyTradeOffQ} proposes a multicast scheduling scheme for Poisson traffic. However, there's no power control and the proposed queue is not always stable\CE. %\cite{CooperativeSch} provides a stochastic optimization for queue lengths in a cache enabled multicast network. However they limit the maximum number of requests that can be served to keep the queue stable. This results in rejection of requests which does not happen in our system. 

Power control in multicast systems has been studied in \cite{Goldsmith, Proakis}. In \cite{Goldsmith}, power allocation optimizes the ergodic capacity 
%(defined as the boundary of the region of all achievable rates in a fading channel with arbitrarily small error probability) 
while maintaining certain minimum rate requirements at the users and average power constraints.   % Authors use water-filling to achieve the optimal policy. 
In \cite{Proakis}, the authors minimize a utility function via linear programming, under SINR constraints at the users and transmit power constraints at the transmitter. Both \cite{Goldsmith,Proakis} derive an optimal power control policy for delivery to all the users, whereas this paper considers delivery to a random subset of users requesting file at that time. %In \cite{Kim}, each packet has a deadline and packets not received by the end of the slot are discarded. The authors use dynamic programming to obtain the optimal policies. 
Also, the power control policies in \cite{Goldsmith, Proakis}, require knowledge of system statistics and are not scalable for our system. Our scheme is computationally scalable, does not require knowledge of system statistics (traffic intensity, fading distributions) and can track changing system statistics.

\textbf{Deep Learning in Wireless Multicast systems:} The ability of DeepRL to handle large state-space dynamic systems is being exploited in various multicast wireless systems/networks. In \cite{Hao}, the authors study a resource allocation problem in unicast and broadcast transmissions. The DeepRL agent learns and selects power and frequency for each channel to improve rate under some latency constraints. Like in our work, they also introduce constraints via Lagrange multipliers. However, the Lagrange multiplier is constant and the agent does not learn it. Thus, the agent also does not adapt if the system dynamics changes as the  Lagrange constant is fixed and the learning rate decays with time. To get the appropriate  Lagrange multiplier is computationally expensive and requires known system statistics. Another work, \cite{Nasir}, applies unconstrained deep reinforcement learning to multiple transmitters for a proportionally fair scheduling policy by adjusting individual transmit powers. %\cite{Dai} applies DeepRL to control power for anti-jamming systems.  
\cite{MulDeepRL} applies DeepRL in queueing in a coded caching based multicast system which is shown to be inferior to our multicast schemes in high traffic rate region. %\cite{VRDeepRL} provides a scheduling strategy in a multicast network for VR Video streaming. Unlike our work, they provide latency reductions for streaming than whole file delivery. 
For more literature on Deep Learning applications to wireless multicast systems, see the detailed survey \cite{MulDeepLSurvey1}.  

%Finally, we also point to some of the existing constrained reinforcement learning algorithms. Model based algorithms (where transition probabilities of the MDP are known) to obtain the optimal policy for 
For Constrained MDPs see \cite{Altman}. However these algorithms are not scalable. %For such systems, several model-free deep (reinforcement learning) algorithms for constrained MDPs are proposed in the literature. %\cite{DBLP:journals/corr/AchiamHTA17, RCPO, SafeCPO}.
\cite{DBLP:journals/corr/AchiamHTA17} introduced constrained reinforcement learning algorithm based on Trust Region Policy optimization. Unlike our case this approach uses discounted constraints. Also, this algorithm requires multiple evaluations of policies and sample paths to reduce the estimator variance. Though this algorithm may perform very well on simulated systems like Atari \cite{DBLP:journals/corr/MnihKSGAWR13}, AlphaGo \cite{Silver1140}, etc., it is not suitable for practical systems where, more often than not, we cannot have multiple evaluations of different policies and sample paths. In \cite{RCPO}, a Lagrange based actor critic approach for constrained RL, is proposed. Since this is also a policy based approach this also suffers from high variance when multiple evaluations are infeasible.           In \cite{SafeCPO}, an alternate approach with two value functions for reward and constraint (cost) with actor-critic policy update, is proposed. Here, at each step a convex relaxation based optimization is used to get the optimal parameter of value functions. We note that the convex optimization step at each iterate is computationally more intensive than a simple SGD step. Thus the above mentioned policy iteration methods either have high variance in practical systems or are computationally intensive. These issues make it difficult to track the changing dynamics in practical systems, as we can in our case. To the best of our knowledge ours is the first constrained value iteration based Deep RL algorithm for constrained MDPs. The use of replay memory and a target network helps reduce estimator variance in our algorithm. These features also increase the practical applicability of our algorithm\CE.             

\indent Rest of the paper is organised as follows. Section \ref{sec:system_model} explains the system model and motivates the problem. Section \ref{DSGD} presents our deep learning based optimal queueing algorithm. Section \ref{sec:PCMQ} motivates the power control problem and briefly explains the power control algorithm proposed in \cite{wcnc}. Section \ref{sec:DRL_power_cont } presents the proposed DeepRL algorithm AC-DQN for scalable, improved power control. Section \ref{sec:IDA} presents our novel deep multi-timescale algorithm to achieve scalable cross-layer optimization of queueing and power control and  provides optimal performance for the multicast system.  Section \ref{sec:simulation} demonstrates our algorithms via simulations and Section \ref{sec:conclusion} concludes the paper. 

\section{System Model} 
\label{sec:system_model}
\begin{figure}[h!]
\centering
\includegraphics[height=5cm,width = 7cm]
{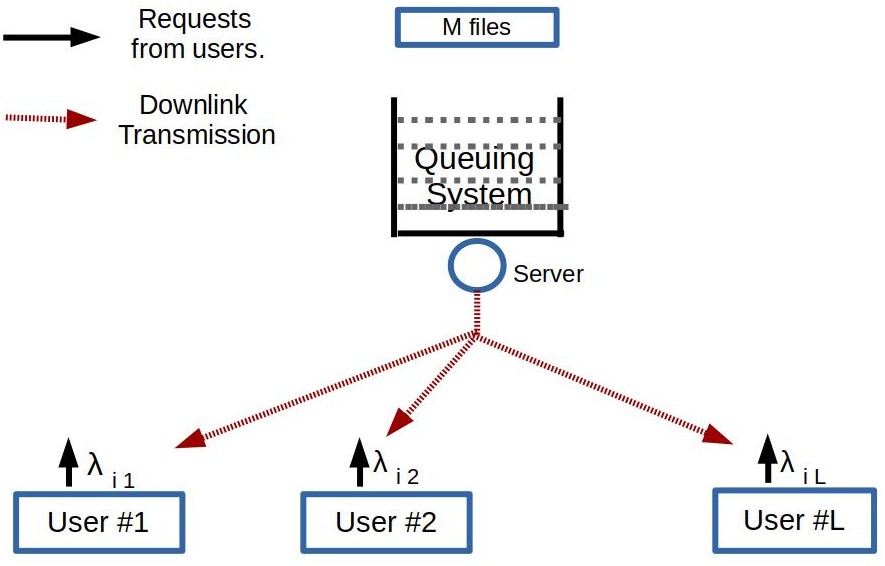}
\caption{System model}
\label{fig:sys_model}
\end{figure}

We consider a system with one server transmitting files from a fixed finite library to a set of users (Figure \ref{fig:sys_model}). We denote the set of users by $\mathcal{L} = \{1,2,\cdots, L\}$ and the set of files by $\mathcal{M} = \{1,2,\cdots, M\}$. %We assume that $M >> L$. 
The request process for file $i$ from user $j$ is a Poisson process of rate $\lambda_{ij}$ which is independent of the request processes of other files from user $j$ and also from other users. The total arrival rate is $\lambda=\sum_{i,j}\lambda_{ij}$. The requests of a file from each user are queued at the server till the user successfully receives the file. All the files are of length $F$ bits. The server transmits at a fixed rate, $R$ bits/sec. Thus, the transmission time for each file is $T = F/R$. 
 
\indent The channels between the server and the users experience time varying fading. The channel gain of each user is assumed to be constant during transmission of a file. The channel gain for the $j^{th}$ user at the $t^{th}$ transmission, is represented by $H_j(t)$. Each $H_j(t)$ takes values in a finite set and form an independent identically distributed (i.i.d) sequence in time, as in \cite{fsmc}. The  channel gains of different users are independent of each other and may have different distributions. Let ${H}=(H_1, \cdots, H_L)$.

\CE Since the requests from the users are queued at the server, every request awaits its turn for transmission and thus experiences a queueing delay which is random in nature. The distribution of this random delay depends on the queueing policy. Also, unsuccessful transmissions due to fading, adds further delay, experienced by each request. We denote by random variable $D$, the overall delay experienced by each request due to both queueing and fading. If $t_A$ is the time of arrival of a request to the server and $t_S$ is time instance representing the end of successful transmission/service of the request. Then the random delay/sojourn time $D$ is given by  $D=t_S-t_R$. Further, $E[D]$ denotes the stationary mean sojourn time experienced by each request.\CE          

 More details of the system are described in the following sections as follows. Section \ref{sec:m_queue} describes the basic Multicast queue proposed in \cite{Arxiv2018}. The queueing schemes to mitigate the effects of fading studied in \cite{wcnc} are also presented. Section \ref{sec:paraQ} parametrizes the queuing schemes. Section \ref{DSGD} provides an online learning scheme to obtain the optimal policy for a given setup.  In Sections \ref{sec:avg_pow_cons} and \ref{sec:MADS}, we summarise  the results from \cite{wcnc}, which show that using power control can further improve the performance and the algorithm used to obtain the optimal power policy.  We will see that this algorithm is not scalable. Then in Section \ref{sec:MDPF}  we provide the MDP of the power control problem. In Section \ref{sec:DRL_power_cont } we will present the scalable  DeepRL  solution for this formulation. 
 %Section \ref{sec:avg_pow_cons} describes the average power constraint and the reward function. Section \ref{sec:MADS} gives the optimization problem and the algorithm considered in \cite{wcnc}. Finally, section \ref{sec:MDPF} gives the MDP formulation of our system.
 
\subsection{Multicast Queue}
\label{sec:m_queue} 
\indent For scheduling transmissions at the server, we consider the \textit{multicast queue} studied in \cite{wcnc}. In this system, the requests for different files from different users are queued in a single queue, called the multicast queue. In this queue, the requests for file $i$ from all users are merged and considered as a single request. The requested file and the users requesting it, is denoted by $(i,\mathbb{L}_i$)\CE. In other words, $\mathbb{L}_i$ is the list of users interested in file $i$\CE. A new request for file $i$, from user $j$ is merged with the corresponding entry $\mathbb{L}_i$, if it already exists. Else, it is appended to the tail of the queue. Service/transmission of file $i$, serves all the users in $\mathbb{L}_i$, possibly with errors due to channel fading.\\
\indent The random subset of users served by the multicast queue at the $t^{th}$ transmission, is denoted by the random binary vector, ${V}(t)=(V_1(t), \cdots, V_L(t))$, where $V_j(t)=1$ implies that the user $j$ has requested the file being transmitted; otherwise, $V_j(t) = 0$. From [Theorem 1, \cite{Arxiv2018}], ${V}(t)$ has a unique stationary distribution. 

\indent It was shown in \cite{Arxiv2018} that the above multicast queue performs much better than the multicast queues proposed in literature before. \textit{The main difference compared to previous multicast schemes is that in this scheme, all requests of all the users for a given file are merged together over time. One direct consequence of this is that the queue length at the base station does not exceed $M$. Thus the delay is bounded for all traffic rates. In fact the mean delays are often better than the coded caching schemes proposed in the literature, as well, for most of the traffic conditions.} 

In a fading scenario, where the different users have independent fading, the performance of this scheme can significantly deteriorate because of multiple retransmissions required to successfully transmit to all the users needed. Thus, in \cite{wcnc}, multiple queuing strategies were proposed and compared to recover the performance of the system and reduce the mean delay substantially. Some of these are also fair to different users in the sense, that the users with good channel gains do not suffer due to users with bad channel gains. We now briefly present the schemes proposed in \cite{Arxiv2018,wcnc} for clarity.
 
\textbf{Retransmit:} This is the simplest scheme proposed in \cite{Arxiv2018}. Here the multicast queue is serviced from head to tail. The head of the line is retransmitted until all the users in it are serviced. The new requests are added to the queue in a similar manner to the simple multicast. This naive scheme works very well in low request rate regime, however performs poorly in the high request rates and severely deteriorates delays experienced by users with good channels.   

\textbf{Single queue with loop-back (1-LB):} The Multicast queue is serviced from head to tail. When a file is transmitted, some of the users will receive the file successfully and some users may receive the file with errors. In the case of unsuccessful reception by some users, the file is retransmitted. A maximum of $N$ $(1 \leq N \leq \infty)$ transmission attempts are made. If there are some users who did not receive the file within $N$ transmission attempts, the request (tuple $(i,\mathbb{L}_i)$ with $\mathbb{L}_i$,   now modified to contain only the set of users who have not received the file $i$ successfully) is fed back to the queue. If there is another pending request in the queue for the same file (a request for the file which came during the current transmission), it is merged with the existing request. Otherwise, a new request for the same file with unsuccessful users is inserted at the tail of the queue.

\textbf{Defer Queue with loop back (2-LB):} This strategy has two queues for servicing the requests. A multicast queue and a defer queue. \CE The multicast queue is similar to the queue mentioned in the beginning of this section and is serviced from head to tail. The defer queue is an additional queue to handle unsuccessful transmissions as follows. \CE When a file is transmitted, some users may receive the file with errors. In the case of unsuccessful reception by some users after a maximum of $N$ transmissions, the file request and the unserviced users are queued in the defer queue. Such requests stay in the defer queue until a new request for the same file arrives. \CE On the arrival of the new request, the new request is merged with the older requests in the defer queue and moved to the tail of the multicast queue. If no such old requests exist in the defer queue, the new request is merged/added to the multicast queue. This queue is shown to provide lower delay to good channel users than to bad channel users.  

Performance of each of these queues, depends on system parameters, transmission power policy, arrival rate, etc. For simplicity of presentation we consider the case of $N=1$ for all the queueing strategies, in this paper.

\subsection{Parametrization of Queueing Strategies}
\label{sec:paraQ}
To adaptively optimize the queueing strategy according to system parameters, it is convenient to first parametrize them. We propose a simple parametrization using probabilities for each queueing strategy. That is, at the end of every service instance if some users did not get the file successfully, the multicast queue chooses to retransmit the head of the line (HoL) request with probability $p_1$, loopback HoL with probability $p_2$ or defer HoL with probability $p_3$, such that $\sum_{j=1}^{j=3} p_j =1$\CE. Thus, $\overline{p}=[p_1,p_2,p_3]$ parametrizes the queueing strategy. Here, $\overline{p}\in \mathbb{P}$, where $\mathbb{P}$ is the probability simplex, $\mathbb{P}=\{[p_1,p_2,p_3]\in [0,1]^3:\sum_{j=1}^{j=3} p_j =1\}$\CE.  Observe, that $\overline{p}=[1,0,0],\ [0,1,0],$ and $[0,0,1]$ represent retransmit, loopback, and defer strategies\CE. In the next section we provide an algorithm to get optimal $\overline{p}$. %We now look at some characteristics of the function we are optimizing, the mean delay experienced by the user and provide an online algorithm to learn the optimum queueing strategy.

\section{Deep Learning for Optimal Queueing}
\label{DSGD}  

	 We are interested in finding the optimal $\overline{p}$ among the parametrized queueing strategies in Section \ref{sec:paraQ} that gives the least average delay. From our previous work (Proposition 1, \cite{Arxiv2018}), it can be shown that for any parameter $\overline{p}$ there exists a stationary mean sojourn time, $E_{\overline{p}}[D]$, where $D$ is the sojourn time and $E$ is the expectation. In this section we propose an online deep learning algorithm to learn $\overline{p}^*=\underset{\overline{p}\in \mathbb{P}}{argmin}E_{\overline{p}}[D]$. However, the map $f:\overline{p}\mapsto E_{\overline{p}}[D]$ is quite complex and it is very difficult to obtain its closed form expression. 
	
%	Since we do not have a closed form expression of $f(\overline{p})$, we depend on its noisy observation $\hat{f}(\overline{p})$ obtained by observing the queueing system, to get $\overline{p}^*$. ReLU (Rectified Linear Unit) based Deep Neural Networks are adept in approximating such complex functions on a compact subsets such as $\mathbb{P}$, \cite{Hanin}. Also, 
%	      
%	
%%	Let, $E_{\overline{p}}[D]$ be the stationary mean sojourn time \CE(mean overall delay experienced by a user from the time its request reach the queue till the user received it successfully), while following the strategy $\overline{p}=[p_1,p_2,p_3]$\CE. From the formulation of parametrization, it is easy to see that\CE:
%\begin{equation}
%\begin{split}
%E_{\overline{p}}[D]=p_1 E_{[1,0,0]}[D]&+p_2 E_{[0,1,0]}[D]\\&+p_3 E_{[0,0,1]}[D],\ \forall \overline{p}\in \mathbb{P}  \end{split}	
%\end{equation}	     
%
%\CE where $D$ is the sojourn time and $E_{\overline{p}}[D]$, is the stationary mean sojourn time for the strategy $\overline{p}$. $E[\cdot]$ is the expectation operator. Clearly, $E_{\overline{p}}$ is a convex function of $\overline{p}$. Thus if, $f({\overline{p}})=E_{\overline{p}}$ is exactly known, in closed form, global minima can be easily obtained using convex optimization methods. 

Since we do not have a closed form expression, we depend on noisy observations of $f$, the mean sojourn time, from the system to get the optimal strategy, $\overline{p}^*$. Here is where Deep Neural Network (DNN) fits in. They are state-of-the-art tools used for several learning problems, especially regression.
	Before we proceed with motivation for using DNN, it is worth mentioning that several stochastic approximation algorithms,  such as simultaneous perturbation stochastic approximation (\cite{Bhatnagar2013SPSA}, pg 41-76), exist for such noisy function optimization. However, convergence of such algorithms are prone to high variance in the gradient estimate and often lead to suboptimal results. In fact we have tried SF-SPSA (\cite{Bhatnagar2013SPSA}, pg 77-102),  in our system and have seen that the algorithm leads to a suboptimal point in many cases. %DNN's on the other hand are known for better generalization in regression problems. It is shown in \cite{Hanin} that the DNNs can approximate any convex function on $[0,1]^d$, $d\in \mathbb{N}$, arbitrarily closely.   
	ReLU (Rectified Linear Unit) based Deep Neural Networks (DNN) on the other hand are adept at approximating such complex functions on compact subsets such as $\mathbb{P}$, \cite{Hanin}. Particularly, it is seen that DNN can provide better generalization in function approximation even with noisy training data \cite{labelnoise}. Further, DNNs are also known to provide good gradient approximates for the approximated function, \cite{DNNGrad}\CE. This motivates us to use DNN to approximate $f(\overline{p})$ as $f_{\theta}(\overline{p})$, where $\theta$ is the weight parameter of the DNN. Further the gradients required for optimization are derived using finite difference method on $f_{\theta}(\overline{p})$. Another important feature of our algorithm is the Replay Memory. This idea is borrowed from the Reinforcement Learning setting \cite{Lin1992}. It helps us in storing previously seen noisy function observations and use it for training the DNN in online fashion. 
	
	The replay memory and online training of the DNN are the important features of our algorithm. Online training inherently adds Importance Sampling \cite{imp_sampl} feature to our algorithm, that is, we train our neural network only with samples that are more informative. This is shown to accelerate DNN training time \cite{imp_sampl}. We will see in our algorithm that this happens naturally as training samples for the neural network come from the parameter $\overline{p}$ update step. These samples give more information about the neighbourhood of the point the algorithm is currently in, thereby improving the confidence/variance in the descent direction. We now present our algorithm, Deep assisted Stochastic Gradient Descent for obtaining the optimal queueing strategy.

\subsection{Deep assisted Stochastic Gradient Descent (DSGD)}

Our algorithm, has three steps:
\begin{itemize}
\item Generating Noisy observation $\hat{f}$ of the function $f$ at random points and storing in replay memory, $\mathbb{M}_D$. This provides us the initial training set. 

\CE		To obtain $\hat{f}$ for a randomly generated point $\overline{p}$, the system is set to follow policy $\overline{p}$ and run till $S_{approx}$ services are completed. Let, $d_i$ be the sojourn time of $i^{th}$ successfully served request in $S_{approx}$ services. These are stored in a temporary memory $\overline{D}$. From $d_i, i\in[\vert\overline{D}\vert]$ compute\CE:
\begin{equation}
\label{eq:noisy_f}
\hat{f}=\frac{1}{\vert\overline{D}\vert}\sum_{i=1}^{\vert\overline{D}\vert} d_i
\end{equation}
The point $(\overline{p},\hat{f})$ is stored in $\mathbb{M}_D$ and $\overline{D}$ is cleared\CE.    
\item Sample a minibatch of points from $\mathbb{M}_D$, uniformly randomly and train $f_{\theta}$:
		\begin{equation}
		\theta\leftarrow \theta - \eta_1 \nabla_{\theta} L_{f_{\theta}}
		\end{equation}			
		where, $L_{f_{\theta}}$ is the Mean Square Error obtaind from minibatch sampled from the replay memory, given by $L_{f_{\theta}}=\sum_{i=1}^n(f_{\theta}(\overline{p}_i)-\hat{f}_i)^2/n$.
\item Obtain numerical gradient of $f_\theta$ at the last executed point $\overline{p}$ and perform a gradient descent: 
		\begin{equation}
		\label{eq:param_step}
		\overline{p}\leftarrow \mathcal{P}(\overline{p} - \eta_2 \nabla_{\overline{p}} {f_{\theta}(\overline{p})})
		\end{equation}
Get the noisy observation of $f$ at the new point. Store the new $(\overline{p},\hat{f})$ to the replay memory, $\mathbb{M}_D$. $\mathcal{P}$ is the projection operator that projects the input to the probability simplex as, 
\begin{equation}
\mathcal{P}[r_1,r_2,r_3]=\{[r_1,r_2,r_3]\}^+/\sum_{i=1}^3\{r_i\}^+)
\end{equation} 
where element wise operator $\{\cdot\}^+=max\{0,\cdot\}$, and $r_i\in \mathbb{R},\ i=1,2,3$.

\item $\eta_1$ and $\eta_2$ are learning parameters and must follow learning rate relationships of multi-timescale stochastic gradient descent, \cite{borkar}, given in (\ref{eq:two_timeline_learning}) in Section \ref{sec:DRL_power_cont }. The detailed algorithm is given in Algorithm \ref{algo:DSGD}.
\end{itemize}

\begin{algorithm}
\SetAlgoLined
\caption{Deep assisted Stochastic Gradient Descent (DSGD) Algorithm}\label{algo:DSGD}
\KwIn{}{Multicast system in \ref{sec:m_queue}, Replay Memory: $\mathbb{M}_D$, Minibatch size: $n$, Training Time: $T_{train}$, Approximation Window: $S_{approx}$, Initialize neural network weights: $\theta$ of $f_{\theta}$, Exploration Parameter: $\epsilon(t)\rightarrow 0$, $\theta,\ \overline{p}$ learning rates: $\eta_1(t)$, $\eta_2(t)$ must satisfy (\ref{eq:two_timeline_learning}), Simulation Time: $T$, Algorithm timeline: $t$, Multicast System timeline: $s$\\}
\For{$t=1$ \KwTo $T$}{
\uIf{($t<T_{train}$)}{	
	$\overline{p}\leftarrow \mathcal{P}(Unif([0,1]^3))$ \\
%	run $S_{approx}$ Multicast services with strategy $\overline{p}$ and store obtained $d_i's$ in $\overline{D}$\CE\\
%	obtain $\hat{f}$, as in (\ref{eq:noisy_f})	$\ \rightarrow\ $
%	clear $\overline{D}$ $\ \rightarrow\ $
%	%$\hat{f}\leftarrow SampleMean(D)$ obtained in the $S_{approx}$ steps.\\
%	store $(\overline{p},\hat{f})$ in $\mathbb{M}_D$\\	
	}
	\Else{
	\textbf{Sample:} Minibatch $n$ from $\mathbb{M}_D$\\
	$/*${Perform DNN $\theta$ update and $\overline{p}$ parameter update as follows:}$*/$\\

	$\theta\leftarrow \theta - \eta_1 \nabla_{\theta} L_{f_{\theta}}$\\
	$\overline{p}\leftarrow \mathcal{P}(\overline{p} - \eta_2 \nabla_{\overline{p}} {f_{\theta}} + Unif([0,\epsilon_t]^3))$\\
%	run $S_{approx}$ Multicast services with strategy $\overline{p}$ and store obtained $d_i's$ in $\overline{D}$\CE\\
%	obtain $\hat{f}$, as in (\ref{eq:noisy_f}) $\ \rightarrow\ $
%	clear $\overline{D}$ $\ \rightarrow\ $
%	store $(\overline{p},\hat{f})$ in $\mathbb{M}_D$\\
	}	
	run $S_{approx}$ Multicast services with strategy $\overline{p}$ and store $d_i's$ in $\overline{D}$\CE\\
	obtain $\hat{f}$, as in (\ref{eq:noisy_f}) $\ \rightarrow\ $
	clear $\overline{D}$\\ %$\ \rightarrow\ $
	store $(\overline{p},\hat{f})$ in $\mathbb{M}_D$\\
}
$\overline{p}^{*}\leftarrow\overline{p}$\\
\KwOut{$\overline{p}^{*}$: Optimal Queueing Strategy}
\end{algorithm}  
We note the following:

\begin{itemize}
\item Initial training phase is necessary to avoid pathological zero gradients in the initial steps, which may inhibit further exploration of the function.
\item The noise exploration in the second SGD step is also for the same reason.
\item Minibatch sampling with Replay memory is to provide IID data samples to the DNN training, which is necessary for better generalization.
\item It is natural to observe that a larger inital training phase and an offline training of DNN may avoid Replay memory during the second SGD learning phase. It is not advisable for the following reasons:

1) At any given point, $\overline{p}$, it is important to have a good estimate (low variance) of the descent direction in (\ref{eq:param_step}). For this it is essential that the DNN approximates the function well in the neighbourhood of $\overline{p}$. In online training this comes naturally, as the consecutive SGD steps in (\ref{eq:param_step}) add more points from this neighbourhood in the replay memory. This is true even if the variance in the gradient estimates are high as the steps in (\ref{eq:param_step}) do not go too far, when the learning rate is choosen appropriately. As the algorithm progresses, more points in neighbourhood are added and the variance in the gradient estimates naturally reduces. This is how our algorithm increases confidence in descent direction using Importance Sampling. To achieve this confidence, with offline training of DNN, it would require prohibitively large training sample set, obtained using Monte Carlo evaluations of $f$.\\      
%\item \CE The online training data generation using replay memory, like in our case, tends to give more points near the minima of the function $f$\CE. This is helpful in better generalization \CE of function $f$ \CE in the neighborhood of the optimal point and in turn improving the gradient estimates.
2) Further with offline training of DNN, the algorithm will not be adaptive if the system dynamics like rates, popularity etc., change. We will see that online training using replay memory is particularly useful \CE when we integrate this algorithm with our power control algorithm.

\item The SGD steps can be replaced with the improvements such as AdamOptimizer \cite{adam} for annealing of the gradients, which helps in stable gradient descent. Infact, we use Adam optimizer in all our SGD steps. 
\end{itemize}

Section \ref{sec:simDSGD} provides the simulation results of DSGD for a multicast system with constant transmit power.  

%\begin{figure}
%\noindent\begin{minipage}{0.5\textwidth}
%\input{DSGD_algo.tex}
%\end{minipage}
%\hspace*{\fill}
%\begin{minipage}{0.5\textwidth}
%\input{ACDQN_algo.tex}
%\end{minipage}
%\end{figure}

\section{Power Control for Multicast Queue}	            
\label{sec:PCMQ}
We now proceed to describe the power control in the Multicast setup. Adapting the transmit power based on system and environment state under certain system constraints helps in providing the power control that may improve QoS, which is quantified by mean user delay under stationarity. It was shown in \cite{wcnc} that  choosing the transmit power based on the channel gains, the system performance improves. We describe the system constraint, a power control model and  the MADS Power control algorithm proposed in \cite{wcnc} in this section. We then end this section with the Markov Decision Process Formulation of the entire system that aids in development of the Deep Reinforcement Learning Based Power Control algorithm.   

\subsection{Average Power Constraint}
\label{sec:avg_pow_cons}
\indent Depending on the value of ${H}(t)$ and ${V}(t)$ at time $t$, the server chooses transmit power $P_t$, based on a power control policy $P_t=\pi({H}(t),{V}(t))$. Choosing a good power control policy is the topic of this section. The state, $S_t$  of the system at time $t$ is $({H}(t), {V}(t))$. Let $\ P_{S_t}$ be the power chosen by a policy for state $S_t$ and $R(S_t, P_{S_t})$ be the number of successful transmissions for the selected power $P_{S_t}$, during the $t^{th}$ service.\\
\indent  For a fixed transmission rate $C$ and for a given  channel gain $H(t)$ of users, the transmit power requirement $P_{req}$ (from Shannon's Formula) for user $j$ is  (assuming file length is long enough) 
\begin{equation}\label{eq:pow_req}
	P_{req}(j,S_t)=\frac{N_g}{H_j^2(t)}(2^{C/B}-1),
\end{equation}	      
where, $B$ is the bandwidth and $N_g$ is the Gaussian noise power at receiver $j$. Here, for simplicity, we are taking the ideal Shannon formula in (\ref{eq:pow_req}), which can be easily modified to make it more realistic (\cite{proakis2008digital}, Chapter 14)\CE. Thus the reward for the chosen power control policy, during $t^{th}$ transmission is given by,
\begin{equation}\label{eq:reward}
	R({S_t},P_{S_t})=\sum_{j=1}^L{V_{j,S_t}\ 1_{\{P_{S_t}>P_{req}(j,S_t)\}}}(t),
\end{equation}
where $V_{j,S_t} = 1$ if the user $j$ has requested the file in service and $V_{j,S_t} = 0$ otherwise. We now describe the Mesh Adaptive Direct Search (MADS) power control policy.
 
\subsection{MADS Power control policy}
\label{sec:MADS}
  The power control policy in \cite{wcnc} is derived from the following optimization problem,
\begin{equation}
\begin{split}
\underset{\{P_1,\cdots,P_K\}}{\max}&{\sum_{k=1}^K{q_k R_k}}\\s.t.\ {\sum_{k=1}^K{q_k P_k}}\leq \overline{P}\ &\text{and}\ P_k\geq 0, k=1,\cdots,K,
\end{split}
\end{equation}
where $\overline{P}$ is the average power constraint, $K$ is the total number of states, $P_k$ is the power chosen by the policy in state $k$, $q_k$ is the stationary distribution of state $k \in \{1,\cdots,K\}$ and are assumed to be known apriori, and $R_k$ is the reward for state $k$, given as $R_k=R(S_t=k,P_t=P_k)$. This is a non-convex optimization problem since the reward in Eq. (\ref{eq:reward}) is a simple function (linear combination of indicators). Mesh Adaptive Direct Search (MADS) \cite{MADS} is used in \cite{wcnc} to solve this constrained optimization problem and obtain the power control policy. Though MADS achieves global optimum, it is not scalable as its computational complexity is very high.

	The state space and action space of this problem can be very high even for a moderate number of users and channel gains, e.g., a system with $L$ users and $G$ channel gain states, has $\mathcal{O}(2^L G^L)$ states. Therefore, in this paper we propose a deep reinforcement learning framework. This not only provides optimal solution for a reasonably large system but does so without knowing the arrival rates and channel gain statistics. In addition, we show via simulations that we can track an optimal solution even when the arrival and channel gain statistics change with time.

\subsection{MDP Formulation}
\label{sec:MDPF}
The above system can be formulated into a finite state, action Markov Decision Process denoted by tuple ($\mathbb{S}, \mathbb{A}, r, \textbf{P}, \gamma$): (state space, action space, reward, transition probability, discount factor), where, transition probability $\textbf{P}(S_{t+1}|S_0,P_0, ..., S_t, P_t)=\textbf{P}(S_{t+1}|S_t, P_t)$, policy $\pi$  chooses power $P_t \sim \pi (.|S_t)$ in state $S_t$ and the instantaneous reward $r_t=R({S_t},P_{t})$.\\  
The action-value function \cite{Puterman:1994:MDP:528623} for this discounted MDP for policy $\pi$ is
\begin{equation}
\begin{split}
Q^{\pi}&(s,a)=\mathbb{E}[\sum_{t=0}^{\infty}{\gamma^t r_t}|S_0=s,P_0=a].
\end{split}
\end{equation}
where $0<\gamma<1$. The optimal $Q\text{-function}$, $Q^*$ is given by $Q^*(s,a)=\underset{\pi}{\max}\ Q^{\pi}(s,a)$ and satisfies the optimality relation,

\begin{equation}\label{eq:bellman}
Q^*(s,a)=r(s,a)+\underset{a'}{\max}\ {\gamma}\mathbb{E}[Q^*(s',a')],
\end{equation}  
where, $s'$ is sampled with distribution $\textbf{P}(.|s,a)$. If we know the optimal Q-function $(Q^{*})$, we can compute the optimal policy via $\pi(s)=\underset{a'}{\arg\max}\ Q^{*}(s,a)$. We know the transition matrix of this system and hence can compute the $Q$-function. But the state space is very large even for a small number of users, rendering the computations infeasible.  Thus, we use a parametric function approximation of the Q function via Deep neural networks and use DeepRL algorithms to get the optimal $Q^{*}$. \CE Our cost function is stationary mean sojourn time.  To get a policy which minimizes this, we actually should be working with average cost MDP instead of discounted MDP.  However, the RL formulation for this problem has been defined for the discounted case, the average case being more complicated. But if we take the discount factor gamma close enough to 1, then the optimal policy obtained via the discounted problem is often close to the average case problem. \CE

Further, to introduce the average power constraint in the MDP formulation, we look at the policies achieving 
\begin{equation}
Q^*(s,a)=\underset{\pi:C_P\leq \overline{P}}{\max}\ Q^{\pi}(s,a)
\end{equation}
where
\begin{equation}\label{eq:long_avg}
C_P=\mathbb{E}[\underset{T\rightarrow\infty}{\lim} \frac{\sum_{t=0}^T P_t}{T}]
\end{equation} 
is the long term average power. We use the Lagrange method for constrained MDPs \cite{Altman} to achieve the optimal policy. In this method, the instantaneous reward is modified as 
\begin{equation}\label{eq:lagrange_cmdp}
	r_t=R(S_t,P_t)-\beta P_t,
\end{equation}    
where, $\beta$ is the Lagrange constant achieving optimal $Q^{*}$ while maintaining, $C_P \leq \overline{P}$. Choosing $\beta$ wrongly will provide the optimal policy with average power constraint different from $\overline{P}$.  
\CE
\section{Deep Reinforcement Learning based Power Control Policy}
\label{sec:DRL_power_cont }
\indent In this section, we describe Deep-Q-Network (DQN) \cite{Mnih2015} based power control. First we describe the DQN algorithm. We then propose a variant of DQN for constrained problems, where in, we use a Lagrange multiplier to take care of the average power constraint. We use multi-time scale stochastic gradient descent approach to also learn the Lagrange multiplier, to obtain the right average power constraint. Finally, we change the learning step size from decreasing to a constant so that the optimal power control can track the time varying system statistics. 
\subsection{Deep Q Networks}
DQN is a popular Deep Reinforcement learning algorithm to handle large state-space MDPs with unknown/complex dynamics, $\textbf{P}(S_{t+1}|S_t, P_t)$. The DQN is a Value Iteration based method, where the action-value function is approximated by a Neural Network. Though there are several follow up works providing improvements over this algorithm \cite{ddpg, ddqn}, we use this algorithm owing to its simplicity. We will show that DQN itself is able to provide us the optimal solution and tracking. These improvements may further improve the performance in terms of sample efficiency, estimator variance etc. 
%The DQN algorithm is given in Algorithm \ref{algo:dqn}. 
Earlier attempts in combining nonlinear function approximators such as neural networks and RL were unsuccessful due to instabilities caused by 1) correlated training samples, 2) drastic change in policy with small change in function approximation, and 3) correlation between the training function and approximated function \cite{DBLP:journals/corr/abs-1810-06339}. Success of DQN is attributed to addressing these issues with two key ingredients of the algorithm: \textbf{Experience Replay Memory} $\mathbb{M}$ and \textbf{Target Network}, $Q_{\theta^{*}}$. The replay memory stores the transitions of an MDP, specifically the tuple, $(S_t, P_t, r_t, S_{t+1})$. The algorithm then samples, uniformly, a random minibatch of transitions from the memory. This removes correlation  between the data and smoothens the data distribution change with iteration. The algorithm has another neural network, approximating the value function, $Q_{\theta}$. The target network and randomly sampled mini-batch from the memory $\mathbb{M}$, form the training set for training the $Q_{\theta}$, at every epoch. This random sampling provides $i.i.d$ samples for performing stochastic gradient descent with loss: %$L^{\pi_{\theta}}_{Q} = \frac{1}{n}\sum_{j=1}^n(Y_j-Q_{\theta}(S_j, A_j))^2$ 
\begin{equation}
L^{\pi_{\theta}}_{Q} = \frac{1}{n}\sum_{j=1}^n(Y_j-Q_{\theta}(S_j, A_j))^2
\end{equation}
 where, $Y_i=r_i+\gamma\ \underset{a'}{\max}Q_{\theta^{*}}(S_i,a'))$. The iterates $\{\theta_t\}$ are given by: 
 %$\theta_{t+1}\leftarrow\theta_t - \eta_1(t) \nabla_{\theta} L^{\pi_{\theta}}_{Q},$
\begin{equation}
\theta_{t+1}\leftarrow\theta_t - \eta_1(t) \nabla_{\theta} L^{\pi_{\theta}}_{Q},
\end{equation}
where $\eta_1(t)$, the step size, satisfies:
\begin{equation}
\label{eq:step_size_dqn}
\sum_{t=0}^{\infty}\eta_1(t)=\infty,\ \sum_{t=0}^{\infty}\eta_1^2(t)<\infty,\ \eta_1(t)\geq 0. 
\end{equation}
The weights of the target network $Q^{*}$ are held constant for $T_{target}$ epochs, thereby controlling any drastic change in policy and reducing correlation between $Q$ and $Q^{*}$. This can be seen as a Risk Minimization problem in nonparametric-regression with {regression function} $Q_{\theta^{*}}$ and {risk} $L^{\pi_{\theta_t}}_{Q}$. Readers are referred to \cite{DBLP:journals/corr/abs-1901-00137} for elaborate analysis of DQN.  Theorem 4.4 in \cite{DBLP:journals/corr/abs-1901-00137} provides a proof of convergence and the rate of convergence using non-parametric regression bounds, when sparse ReLU networks are used, under certain smoothness assumptions on the reward function and the dynamics.

\subsection{Adaptive Constrained DQN (AC-DQN)}
The DQN algorithm is meant for unconstrained optimization. Since our problem has an average power constraint of $\overline{P}$, we consider the instantaneous reward in (\ref{eq:lagrange_cmdp}), with a Lagrange multiplier $\beta$. The long term constraint depends on the Lagrange multiplier and can be quite sensitive to it. Thus, we design our algorithm, AC-DQN, to learn the appropriate $\beta$. We will see later, that this will enable us to further modify our algorithm to track the changing statistics of the  channel gains and arrival statistics. The AC-DQN algorithm is given in Algorithm \ref{algo:AC-DQN}. Here, we use multi-timescale SGD as in \cite{borkar}. In this approach, in addition to the SGD on $Q_{\theta}$, using minibatch, we use a stochastic gradient descent on the Lagrange constant, $\beta$ as 
%$\beta_{t+1} \leftarrow \beta_{t} - \eta_2(t) \nabla_{\beta} L_P^{{\pi}_{\theta}}$ 
\begin{equation}
\beta_{t+1} \leftarrow \beta_{t} + \eta_2(t) \nabla_{\beta} L_P^{{\pi}_{\theta}}, 
\end{equation}
where $\nabla_{\beta} L_P^{{\pi}_{\theta}} = C_P(S_t)-\overline{P}$. Since the expectation in (\ref{eq:long_avg}) is not available to us, we take $C_P(S_t)={\sum_{i=t-T_W}^t P_i(S_i)}/{T_W}$, where $T_W$ is the finite horizon window.
%\begin{equation}
%\nabla_{\beta} L_P^{{\pi}_{\theta}} = C_P(S)-\overline{P}
%\end{equation}
Additionally $\eta_1$ and $\eta_2$ are required to follow \cite{borkar}:
\begin{equation}
\begin{split}
\label{eq:two_timeline_learning}
\sum_{i=1}^{\infty}\eta_1(i)&=\sum_{i=1}^{\infty} \eta_2(i)=\infty,\\ \sum_{i=1}^{\infty}\eta_1^2(i)+\eta_2^2(i)&<{\infty},\ \frac{\eta_2(i)}{\eta_1(i)}\rightarrow 0.
\end{split}
\end{equation}        

\begin{algorithm}
\SetAlgoLined
\caption{Adaptive Constrained DQN (AC-DQN) Algorithm}\label{algo:AC-DQN}
\KwIn{}{MDP-$(\mathbb{S},\mathbb{A}, r, \textbf{P}, \gamma)$, $r$ as in (\ref{eq:lagrange_cmdp}), Replay Memory: $\mathbb{M}$, Minibatch size: $n$, Initialize $T$, $T_{target}$, $\theta, \theta^{*}$ of $Q_{\theta}$ and $Q_{\theta^{*}}$, Exploration Parameter: $\epsilon(t)\rightarrow 0$, Lagrange Constant: $\beta$, Value and Lagrange learning rates: $\eta_1(t)$,\ $\eta_2(t)$ must satisfy (\ref{eq:two_timeline_learning}), Initialize $T_W$\\}
\For{$t=1$ \KwTo $T$}{
	Observe state $S_t$,	Apply action $A_t=\pi_t(S_t)=\underset{a'}{\arg\max\ }{Q_{\theta}(S_t,a')}$, $\epsilon$-greedily\\
	Observe: $r_t,S_{t+1}$ \\ Store: $(S_t, A_t, r_t, C_P(S_t), S_{t+1})$ in $\mathbb{M}$\\
	\textbf{Sample:} Minibatch $n$ from $\mathbb{M}$\\
	\For{$i=1$ \KwTo $n$}{
	$Y_i=r_i+\gamma\ \underset{a'}{\max}Q_{\theta^{*}}(S_{i+1},a')$
%	Train $Q_{\theta}$ using ${Y_i}$ using SGD\\
	}
$/*${Perform two time-scale stochastic gradient descent as follows:}$*/$\\	
$\theta\leftarrow \theta - \eta_1 \nabla_{\theta} L^{\pi_{\theta}}_{Q}$\\
$\beta \leftarrow \beta + \eta_2 \nabla_{\beta}L^{\pi_{\theta}}_{P}$ \\
	at every ${t=m T_{target}}, {m\in \mathbb{N^{+}}}$: update $\theta^{*}\leftarrow{\theta}$
}
$\pi^{*}\leftarrow\pi_T$, $\theta^{*}\leftarrow{\theta}$\\
\KwOut{$Q_{\theta^{*}}$: Optimal $Q$-Function, $\pi$: Optimal Policy}
\end{algorithm}  

%\begin{algorithm}
%\SetAlgoLined
%\caption{Adaptive Power Control DQN (AC-DQN) Algorithm}\label{algo:AC-DQN}
%\KwIn{MDP-$(\mathbb{S},\mathbb{A}, r, \textbf{P}, \gamma)$, $r$ as in (\ref{eq:lagrange_cmdp}), Replay Memory: $\mathbb{M}$, Minibatch size: $n$, Initialize $T$, $T_{target}$, Initialize weights $\theta, \theta^{*}$ of $Q_{\theta}$ and $Q_{\theta^{*}}$, Exploration Parameter: $\epsilon(t)\rightarrow 0$, Lagrange Constant: $\beta$, Value learning rate $\eta_1(t)\rightarrow 0$, Lagrange learning rate: $\eta_2(t)\rightarrow 0$ satisfying (\ref{eq:two_timeline_learning}), Initialize $T_W$}
%\For{$t=1$ \KwTo $T$}{
%	Observe state $S_t$,	Apply action $A_t=\pi_t(S_t)=\underset{a'}{\arg\max\ }{Q_{\theta}(S_t,a')}$, $\epsilon$-greedily\\
%	Observe: $r_t,S_{t+1}$ \\ Store: $(S_t, A_t, r_t, C_P(S_t), S_{t+1})$ in $\mathbb{M}$\\
%	\textbf{Sample:} Minibatch $n$ from $\mathbb{M}$\\
%	\For{$i=1$ \KwTo $n$}{
%	$Y_i=r_i+\gamma\ \underset{a'}{\max}Q_{\theta^{*}}(S_{i+1},a')$
%%	Train $Q_{\theta}$ using ${Y_i}$ using SGD\\
%	}
%$/*${Perform two time-scale stochastic gradient descent as follows:}$*/$\\	
%$\theta\leftarrow \theta - \eta_1 \nabla_{\theta} L^{\pi_{\theta}}_{Q}$\\
%$\beta \leftarrow \beta + \eta_2 \nabla_{\beta}L^{\pi_{\theta}}_{P}$ \\
%	at every ${t=m T_{target}}, {m\in \mathbb{N^{+}}}$: update $\theta^{*}\leftarrow{\theta}$
%}
%$\pi^{*}\leftarrow\pi_T$, $\theta^{*}\leftarrow{\theta}$\\
%\KwOut{$Q_{\theta^{*}}$: Optimal $Q$-Function, $\pi$: Optimal Policy}
%\end{algorithm}  

\textbf{Tracking with AC-DQN:} Tracking of system statistics is essential, to achieve optimal power control in a non-stationary system. In multi-time scale stochastic gradient descent, such as AC-DQN, step sizes $\eta_1(t)$ and $\eta_2(t)$ can be fixed to enable tracking. If $\eta_2<<\eta_1$, then the Lagrange multiplier changes much more slowly than the $Q$-function. Then the two timescale theory (see, e.g., \cite{borkar}), will allow the Lagrange multiplier to adapt slowly to the changing system statistics but at the same time provide average power control. The solution will reach in a neighbourhood of the optimal point. Although the convergence of this modified algorithm is not proved yet (even for the unconstrained DQN, convergence has been proved only recently  in \cite{DBLP:journals/corr/abs-1901-00137}), our simulations will show that the resulting algorithm tracks the optimal solution in the time varying scenario.

The time varying scenario in our setup results due to change in the request arrival statistics from the users and changing channel gain statistics due to motion of the users.

\section{Integrated DSGD and AC-DQN (IDA)}
\label{sec:IDA}
	We are now familiar with how multi-time scale stochastic gradient descent can be used for optimization of a stochastic system with multiple objectives. We extend this idea to learn the optimal queueing strategy while learning the optimal power control policy and simultaneously satisfying the average power constraint. Towards this we add DSGD as a third timescale to AC-DQN. Though DSGD internally has two stochastic gradient descent steps, we consider it as a combined third step of IDA for conceptual clarity. We present our Integrated DSGD and AC-DQN (IDA) in Algorithm \ref{algo:IDA}. There are four learning rates involved in the algorithm. The four learning rates are supposed to satisfy the following criteria for convergence of the algorithm \cite{borkar}: %\CE verifying if this is the sufficient condition\CE        
\begin{equation}
\label{eq:multi_timeline_learning}
\begin{split}
\sum_{i=1}^{\infty}\eta_j(i)=\infty,&\ j=1,2,3,4,\\ \sum_{i=1}^{\infty}\sum_{j=1}^{4}\eta_j^2(i)<{\infty},&\ \frac{\eta_{j+1}(i)}{\eta_j(i)}\rightarrow 0,\ j=1,2,3.
\end{split}
\end{equation}        

Though this criterion is required for convergence, we have seen that constant step sizes are helpful in tracking. So we will see our simulations with $\eta_1>\eta_2>\eta_3/T_{approx}>\eta_4/T_{approx}$.\\  

\begin{algorithm}
%\setstretch{1.0}
\SetAlgoLined
\caption{Integrated DSGD and AC-DQN Algorithm (IDA)}\label{algo:IDA}
\KwIn{}{\textbf{DQN Input}: MDP-$(\mathbb{S},\mathbb{A}, r, \textbf{P}, \gamma)$, $r$ as in (\ref{eq:lagrange_cmdp}), $\mathbb{M}$, $n$, $T$, $T_{target}$, $\theta, \theta^{*}$ of $Q_{\theta}$, $Q_{\theta^{*}}$, $\epsilon(t)\rightarrow 0$, $\beta$, $\eta_1(t)\rightarrow 0$, $\eta_2(t)\rightarrow 0$ and $T_W$ are same as in Algorithm \ref{algo:AC-DQN},\\ 
\textbf{DSGD Inputs:} Replay Memory: $\mathbb{M}_D$, Minibatch size: $n_D$, Training Time: $S_{train}$, Approximation Window: $T_{approx}$, Initialize weights $\theta^{\dagger}$ of $f_{\theta^{\dagger}}$, Exploration Parameter: $\epsilon_D(s)\rightarrow 0$, Learning rates: $\eta_i, i=1,2,3,4$ satisfy,  (\ref{eq:multi_timeline_learning}), System timeline: $t$, DSGD timeline: $s$,}
$s\leftarrow 0$, $\overline{p}\leftarrow \mathcal{P}(Unif([0,1]^3))$\\
\For{$t=1$ \KwTo $T$}{
%	Observe state $S_t$,	Apply action $A_t=\pi_t(S_t)=\underset{a'}{\arg\max\ }{Q_{\theta}(S_t,a')}$, $\epsilon$-greedily\\
%	Observe: $r_t,S_{t+1}$ $\ \rightarrow\ $ Store: $(S_t, A_t, r_t, C_P(S_t), S_{t+1})$ in $\mathbb{M}$\\
	Observe $S_t$, Take action $A_t$ and store $(S_t, A_t, r_t, C_P(S_t), S_{t+1})$ in $\mathbb{M}$\\
	\textbf{Sample:} Minibatch $n$ from $\mathbb{M}$ as in Algorithm \ref{algo:AC-DQN}\\
%	\For{$i=1$ \KwTo $n$}{
%	$Y_i=r_i+\gamma\ \underset{a'}{\max}Q_{\theta^{*}}(S_{i+1},a')$
%%	Train $Q_{\theta}$ using ${Y_i}$ using SGD\\
%	}
$/*${Perform two time-scale stochastic gradient descent as follows:}$*/$\\	
$\theta\leftarrow \theta - \eta_1 \nabla_{\theta} L^{\pi_{\theta}}_{Q}$,\\$\beta \leftarrow \beta + \eta_2 \nabla_{\beta}L^{\pi_{\theta}}_{P}$ \\
	at every ${t=m T_{target}}, {m\in \mathbb{N^{+}}}$: update $\theta^{*}\leftarrow{\theta}$\\
	
$\overline{D}\leftarrow append(sojourtime\ d_i's)$ in current service	\\

\uIf{$t=m T_{approx}, {m\in \mathbb{N^{+}}}$}{
\uIf{($s<S_{train}$)}{	
	$\overline{p}\leftarrow \mathcal{P}(Unif([0,1]^3))$ \\
%\CE	obtain $\hat{f}$, as in (\ref{eq:noisy_f})	$\ \rightarrow\ $
%	clear $\overline{D}$ $\ \rightarrow\ $	
%	store $(\overline{p},\hat{f})$ in $\mathbb{M}_D$\\	
	}
	\Else{
	\textbf{Sample:} Minibatch $n_D$ from $\mathbb{M}_D$\\
	$/*${Perform DNN $\theta^{\dagger}$ update and $\overline{p}$ parameter update as follows:}$*/$\\

	$\theta^{\dagger}\leftarrow \theta^{\dagger} - \eta_3 \nabla_{\theta^{\dagger}} L_{f_{\theta_{\dagger}}}$,\\
	$\overline{p}\leftarrow \mathcal{P}(\overline{p} - \eta_4 \nabla_{\overline{p}} {f_{\theta^{\dagger}}(\overline{p})} + Unif([0,\epsilon_D(s)]^3))$\\
%	\CE	obtain $\hat{f}$, as in (\ref{eq:noisy_f})	$\ \rightarrow\ $
%	clear $\overline{D}$ $\ \rightarrow\ $
%	store $(\overline{p},\hat{f})$ in $\mathbb{M}_D$\\
	}
	\CE	obtain $\hat{f}$, as in (\ref{eq:noisy_f})\\%	$\ \rightarrow\ $
	clear $\overline{D}$\\ %$\ \rightarrow\ $
	store $(\overline{p},\hat{f})$ in $\mathbb{M}_D$\\
	$s\leftarrow s+1$\\
	%clear($\overline{D}$)\\
}
}
$\overline{p}^{*}\leftarrow\overline{p}$, $\pi^{*}\leftarrow\pi_T$, $\theta^{*}\leftarrow{\theta}$\\
\KwOut{$\overline{p}^{*}$: Optimal Queueing Strategy, $Q_{\theta^{*}}$: Optimal $Q$-Function, $\pi$: Optimal Policy}
\end{algorithm}  

We note the following:
\noindent\begin{itemize}
\item This is a generalized algorithm that can be used in systems where multiple objectives are to be met simultaneously such as in cross-layer designs in wireless networks.
\item In applying multi-time scale stochastic optimization, it is necessary to identify which parameters are to be learnt in a faster timescale and which in slower. In IDA we learn the queueing strategy in a slower time scale and the power control policy on a faster timescale. %This is important for stability reasons. 
To have a meaningful update of Q-function network $\theta$, it is necessary that the underlying MDP doesn't change drastically. This is ensured by updating the ${\theta^{\dagger}}$ and queueing strategy $\overline{p}$ at a much slower rate as compared to the $\theta$ and $\theta^{*}$ updates.
\item In practical systems, where the systems statistics are usually non-stationary, the learning rate selection ($\eta_i,\ i=1,2,3,4$ selection) is the most important engineering decision. This controls the trade-off between speed and stability of the algorithm. Learning rates must be carefully selected to ensure that the parameter updates are neither too slow to track the changing system statistics nor too fast for stability of the learning algorithm.  
\end{itemize}

We now present simulation results of all the algorithms presented in this paper. \CE 
\section{Simulation Results and Discussion}
\label{sec:simulation}
\indent In this section, we first present simulation results for our DSGD algorithm. We run the multicast system with constant transmit power. We compare the performance of our DSGD queueing algorithm against each queueing strategy proposed in \cite{wcnc}. Next, we demonstrate the Deep Learning methods for power control proposed in this paper. We compare the performances of AC-DQN and MADS Power control policies. Though MADS provides optimal solutions for small system sizes, it is not scalable. We show that the Deep Learning algorithm, AC-DQN, indeed achieves the global optimum obtained by MADS algorithm, while being scalable with the system size (number of users). We further demonstrate that AC-DQN algorithm tracks the changing system dynamics and obtains the optimal policy, adaptively. Finally, we present our integrated algorithm for optimal queuing and power control, the IDA algorithm. We show, numerically, that the algorithm achieves the optimal point obtained by both DSGD and AC-DQN. Our multicast system is implemented in Python and we use Keras libraries in Python for implementation of our algorithms\footnote{The system and algorithm codes are available in https://github.com/rkraghu88/SchedulingPC$\_$IDA}.

%We consider three systems, first one with 10 users to demonstrate DSGD performance. Second, with 4 users to compare AC-DQN and MADS and third with 20 users, showing performance of AC-DQN. MADS is not able to provide a solution for the third system since the space and time complexity of MADS increase exponentially with the number of users. In MADS and AC-DQN examples, we split the users in two equal sized groups, one group has good channel statistics and the other bad channel statistics, to show the advantage of power control. In both the systems, we compare all the algorithms with a constant power control policy, where the transmit power $P_t$ is fixed to $P_t=\overline{P}$, to indicate the gain due to power control. We use the 10 user system to demonstrate the IDA algorithm. We don't show the performance of good and bad channel users individually, rather provide the overall mean sojourn time. This is just to add clarity to the IDA concept. The system and algorithm parameters, used for the simulations are as follows:  

\subsection{Simulation parameters}
We consider three systems with varying system configurations as follows: 
%\vspace{-20pt}
\subsubsection{\textbf{Small User Case}}\label{sec:system_4} Number of users, $L=4$, Catalog Size $M= 100$, File Size $F=10MB$, Transmission rate $C=10MB/s$, Bandwidth $B=10MHz$, Channel Gains $\sim$ Uniform([0.1 0.2 0.3]) for two users with bad channel statistics and $\sim$ Uniform([0.7 0.8 0.9]) for two users with good channel statistics, File Popularity: Uniform, (Zipf exponent = 0), Average Power Constraint  $\overline{P}=7$, Simulation time= $10^5$ mutlicast transmissions. 
%\vspace{-20pt}
\subsubsection{\textbf{Moderate User Case}}\label{sec:system_10} System Parameters: Power Transmit Levels = 20 (1 to 50), $L=10$, $M = 100$, $F=10MB$, $C=10MB/s$, Channel Gains: Exponentially distributed ($\sim exp(0.1)$ for bad channel, $\sim exp(1.0)$ for good channel), $R=10MB/s$, $B=10MHz$, $\overline{P}=7$, File Popularity: Zipf distribution with Zipf exponent = 1. Simulation time: $10^5$ multicast transmissions. In both the cases, we set the noise power as $N_g=1$.
\subsubsection{\textbf{Large User Case}}\label{sec:system_20} System Parameters: Same as \ref{sec:system_10} except, $L=20$. %Power Transmit Levels = 20 (1 to 50), $L=20$, $M = 100$, $F=10MB$, $C=10MB/s$, Channel Gains: Exponentially distributed ($\sim exp(0.1)$ for bad channel, $\sim exp(1.0)$ for good channel), $R=10MB/s$, $B=10MHz$, $\overline{P}=7$, File Popularity: Zipf distribution with Zipf exponent = 1. Simulation time: $10^5$ multicast transmissions. In all the cases, we set the noise power as $N_g=1$.
%\CE We use the following hyperparameters for our algorithms:\CE
%\vspace{-20pt}
\subsubsection{\textbf{Hyperparameters}}

For DSGD, we consider a fully connected neural network with two hidden layers. First layer has 32 nodes and second layer has 16 nodes. All layers have ReLU activation function. $\mathbb{M}_D=1000$, Minibatch size: $n_D=50$, $T_{train}=100$: Training Time, $S_{approx}=100$: Approximation Window, Initialize weights $\theta$ of $f_{\theta}$, $\epsilon(t)\rightarrow 0$, $\eta_1(t)=.01/(1+.00001 t)$, $\eta_2(t)=.001/(1+.00001 t log (log (t)))$.

In AC-DQN, we consider fully connected neural networks  with two hidden layers for all the function approximations considered in the algorithms.   Input layer nodes are assumed to be $2L$ and the output layer nodes is equal to 20, the number of transmit power levels. Each output represents the Q value for a particular action. The action space is restricted to be finite, as DQN converges only with finite action spaces.  We use two hidden layers for the neural network, with 128 and 64 nodes, and ReLU activation function is chosen, respectively. The other parameters are as follows: Replay memory size $\vert\mathbb{M}\vert=30000$,
$\gamma= 0.9$, $\epsilon_0=1.0$, $\epsilon_{decay}= 0.98$, $\epsilon_t = \epsilon_0 (0.98)^t$, $\eta_1= 0.001$, $\eta_1^{decay}=0.00001$, $\eta_2=.0001$, $\eta_2^{decay}=0.00001$, Mini-batch Size $(n)=64$, $T_{target}=100$, and $T_W=200$.

Finally, in IDA algorithm we combine the parameters of both DSGD and AC-DQN. Step sizes are however held constant with value of each step size at $t=0$.

\subsection{Optimal Queueing using DSGD}
\label{sec:simDSGD}
\CE We consider the moderate user system in section \ref{sec:system_10} for demonstrating the performance of DSGD. %As used widely in studies concerning Content Centric Networks such as ours, 
We assume the widely accepted IRM traffic model, with unity zipf popularity for the 100 different file requests arriving at 10 users\CE. The server is endowed, in different simulation runs, with different queueing strategies. We compare our DSGD based queueing strategy at server with the individual queueing strategies, mentioned in section \ref{sec:system_model}. The server transmits the files with constant transmit power $\overline{P}=7$. We model the wireless fading to follow Rayleigh distribution. This introduces the errors in file transmissions\CE. 

\begin{figure}
\noindent\begin{subfigure}{0.45\textwidth}
%\hspace{-20pt}
\centering
% This file was created by matlab2tikz.
%
%The latest updates can be retrieved from
%  http://www.mathworks.com/matlabcentral/fileexchange/22022-matlab2tikz-matlab2tikz
%where you can also make suggestions and rate matlab2tikz.
%
\definecolor{mycolor1}{rgb}{0.00000,0.44700,0.74100}%
\definecolor{mycolor2}{rgb}{0.85000,0.32500,0.09800}%
\definecolor{mycolor3}{rgb}{1.00000,0.00000,1.00000}%
\begin{tikzpicture}

\begin{axis}[%
width=2.7in,
height=1.5in,
at={(0in,0in)},
scale only axis,
xmin=0.199999999999989,
xmax=3,
xlabel={Total Arrival Rate},
ymin=0,
ymax=400,
ylabel={Mean Sojourn Time (Sec)},
axis background/.style={fill=white},
xmajorgrids,
ymajorgrids,
legend style={nodes={scale=0.7, transform shape}, at={(0.237,0.696)}, anchor=south west, legend cell align=left, align=left, draw=white!15!black}
]
\addplot [color=mycolor1, dashed, line width=2.0pt, mark size=3.0pt, mark=+, mark options={solid, mycolor1}]
  table[row sep=crcr]{%
0.199999999999989	6.22696034150965\\
0.399999999999977	18.9746379330613\\
0.600000000000023	58.3464790420553\\
0.800000000000011	104.169462060438\\
1	145.480547473121\\
2	264.486139781896\\
3	352.677126146442\\
};
\addlegendentry{Retransmit}

\addplot [color=mycolor2, dashdotted, line width=2.0pt, mark size=3.0pt, mark=x, mark options={solid, mycolor2}]
  table[row sep=crcr]{%
0.199999999999989	5.58546186831265\\
0.400000000000006	11.9959726036882\\
0.599999999999994	45.2748104549452\\
0.800000000000011	64\\
1	98.3085834722048\\
2	174.51496687787\\
3	192.521291509862\\
};
\addlegendentry{Loopback}

\addplot [color=mycolor3, dotted, line width=2.0pt, mark=o, mark options={solid, mycolor3}]
  table[row sep=crcr]{%
0.199999999999989	79.3746624772538\\
0.400000000000006	96.8718127846145\\
0.599999999999994	115.943357734115\\
0.800000000000011	135.571444970381\\
1	145.108819648405\\
2	139.005002261183\\
3	144.038254803095\\
};
\addlegendentry{Defer}

\addplot [color=black, line width=2.0pt, mark=diamond, mark options={solid, black}]
  table[row sep=crcr]{%
0.199999999999989	5.63106091774222\\
0.400000000000006	16.3379737121734\\
0.599999999999994	45.1352341648988\\
0.800000000000011	66.3500332445363\\
1	104.636989941756\\
2	143.211501208732\\
3	139.383527044306\\
};
\addlegendentry{DSGD}

\end{axis}
\end{tikzpicture}%
\caption{DSGD Mean Sojourn Times vs Arrival Rate.}
\label{fig:dsgd_cp}
\end{subfigure}

\begin{subfigure}{0.45\textwidth}
\vspace{20pt}
\centering
\input{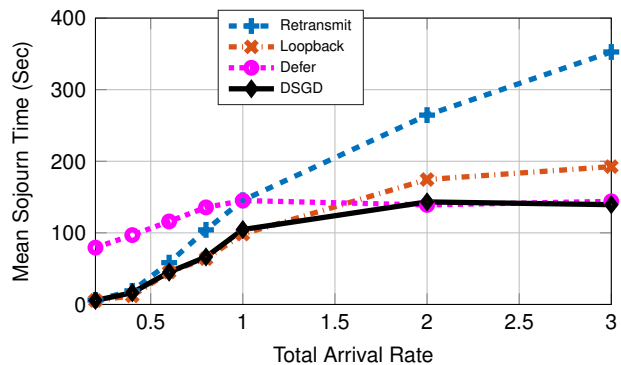}
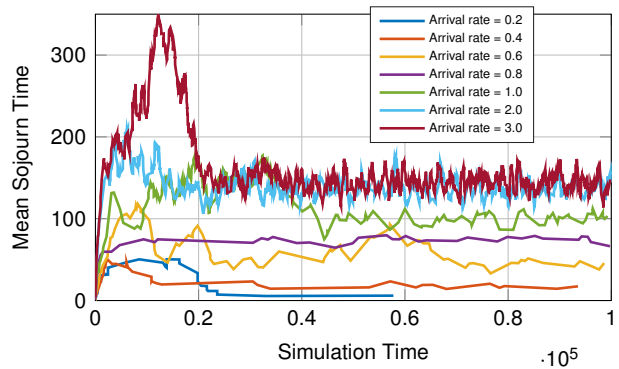
\caption{DSGD Convergence of Mean Sojourn Time.}
\label{fig:dsgd_conv}
\end{subfigure}

\vspace{20pt}
\caption{DSGD Performance in parametrized multicast system with constant power policy, $L=10$, $\overline{P}=7$, Zipf Popularity (Zipf exponent =1), Rayleigh fading with mean, 0.1 and 1.0 for bad and good users respectively.}
\end{figure}

\begin{figure}
%\centering
%\includegraphics[trim={3.5cm 9cm 3.5cm 9.5cm},clip,height=4cm,width=8.5cm]{JournalFigures/Prob_p6.pdf}
\begin{subfigure}{0.45\textwidth}
% This file was created by matlab2tikz.
%
%The latest updates can be retrieved from
%  http://www.mathworks.com/matlabcentral/fileexchange/22022-matlab2tikz-matlab2tikz
%where you can also make suggestions and rate matlab2tikz.
%
\definecolor{mycolor1}{rgb}{0.00000,0.44700,0.74100}%
\definecolor{mycolor2}{rgb}{0.85000,0.32500,0.09800}%
\definecolor{mycolor3}{rgb}{0.92900,0.69400,0.12500}%
\begin{tikzpicture}

\begin{axis}[%
width=2.7in,
height=1.2in,
at={(0in,0in)},
scale only axis,
xmin=0,
xmax=100000,
xlabel={Simulation Time},
ymin=0,
ymax=1,
ylabel={Porbability Vector},
axis background/.style={fill=white},
xmajorgrids,
ymajorgrids,
legend style={nodes={scale=0.7, transform shape}, at={(0.584,0.758)}, anchor=south west, legend cell align=left, align=left, draw=white!15!black}
]
\addplot [color=mycolor1, dashdotted, line width=1.5pt]
  table[row sep=crcr]{%
0	0.540291943223565\\
1200	0.60103380313376\\
3000	0.629755241650855\\
3700	0.613262280414347\\
4800	0.558008672742289\\
6500	0.473389281731215\\
7200	0.457630203134613\\
8300	0.441792083962355\\
8800	0.406168465648079\\
9300	0.35167993781215\\
10000	0.252022616579779\\
11300	0.0602286490029655\\
11800	0\\
16900	0.00882721805828623\\
17900	0.0433772868273081\\
18800	0.0908359662862495\\
19600	0.151712889855844\\
20800	0.259720560337882\\
21900	0.383183809375623\\
22100	0.401440817120601\\
22300	0.390338337223511\\
22700	0.336578467249637\\
23000	0.316808019837481\\
23400	0.312176167353755\\
23900	0.329461687928415\\
24400	0.364980612546788\\
24700	0.38496933937131\\
25300	0.385065107038827\\
25800	0.342415776991402\\
26200	0.332319210399874\\
26700	0.342361573289963\\
28000	0.387032746250043\\
28300	0.364141101104906\\
28700	0.345936966477893\\
29200	0.34473272023024\\
29700	0.359506122447783\\
30200	0.376646735254326\\
30800	0.364637137143291\\
32800	0.360529854020569\\
33300	0.361162584493286\\
34500	0.370569441380212\\
35000	0.364712938302546\\
35800	0.379712038018624\\
36800	0.404073534547933\\
37200	0.382232790347189\\
37700	0.37795019026089\\
38500	0.37915170515771\\
38700	0.363438807878993\\
39100	0.316448496509111\\
39500	0.292845723524806\\
40000	0.283087474585045\\
40700	0.291528016823577\\
41500	0.319302632429753\\
42900	0.385253234926495\\
43500	0.36033252395282\\
44000	0.364733649388654\\
44900	0.385834798697033\\
45100	0.364797222136986\\
45400	0.326212340471102\\
45800	0.296011798665859\\
46300	0.278925832957611\\
46900	0.276207079587039\\
47800	0.294278882312938\\
49300	0.342970893369056\\
51000	0.413082329047029\\
52600	0.350288984263898\\
53300	0.359048404920031\\
54300	0.371380254087853\\
55400	0.37567348095763\\
55800	0.381532833358506\\
56500	0.361736569000641\\
57200	0.363732311729109\\
58000	0.387730365619063\\
58400	0.395751964708325\\
58800	0.358206890581641\\
59200	0.331654831650667\\
59700	0.317670302611077\\
60300	0.320325449749362\\
61600	0.349391043913784\\
62800	0.37656381171837\\
64300	0.383975860851933\\
65200	0.341103834332898\\
66100	0.319601371869794\\
66900	0.32240914387512\\
67600	0.324509708909318\\
68200	0.318225368260755\\
70600	0.322806463678717\\
71700	0.278390777079039\\
72200	0.272487898444524\\
73000	0.286940776291885\\
74000	0.3085081087047\\
76000	0.278259650076507\\
77900	0.298073947415105\\
79100	0.319776194883161\\
80900	0.305750853673089\\
81400	0.308762730564922\\
82500	0.291884977938025\\
83100	0.305689879402053\\
83900	0.327268605789868\\
84400	0.311169539796538\\
85000	0.286698888623505\\
85500	0.275119530691882\\
86300	0.242834769582259\\
87000	0.245923169291927\\
87600	0.266602361196419\\
88200	0.28765654702147\\
89600	0.322217264096253\\
90300	0.297225483867805\\
90900	0.283778251134208\\
91700	0.290677697063074\\
92500	0.294081366970204\\
93500	0.287028817416285\\
95000	0.310994526793365\\
96500	0.278225883303094\\
96600	0.279668789677089\\
};
\addlegendentry{Retransmit}

\addplot [color=mycolor2, dotted, line width=1.5pt]
  table[row sep=crcr]{%
0	0.0666827747918433\\
1000	0\\
2600	0.00530309835448861\\
3100	0.033491323236376\\
3700	0.0895076210581465\\
4400	0.180388875422068\\
6400	0.446366343196132\\
6900	0.484403254784411\\
7500	0.512699208702543\\
9200	0.572539044253062\\
9800	0.623550653632265\\
11900	0.818121856456855\\
13000	0.821097441847087\\
13900	0.83987494847679\\
15500	0.89846366585698\\
16900	0.98236085670942\\
17000	0.988594304726576\\
18100	0.947641412451048\\
19000	0.895677655134932\\
20000	0.814465929521248\\
21000	0.719272854476003\\
22200	0.599737236494548\\
23500	0.628558758849977\\
24400	0.618609528901288\\
25200	0.608671046706149\\
25700	0.622722466549021\\
26500	0.632421127869748\\
27400	0.625601439693128\\
28000	0.611456398386508\\
29100	0.632933449436678\\
29900	0.631350567608024\\
30300	0.625820936984383\\
30900	0.635119654616574\\
32600	0.630749534946517\\
33300	0.638837415506714\\
34800	0.627841715235263\\
35700	0.622645878975163\\
37000	0.597695829841541\\
37800	0.599752541806083\\
38500	0.602349907741882\\
40100	0.632957980007632\\
41400	0.620600440044655\\
43200	0.599096837264369\\
43900	0.600063755438896\\
45100	0.592287961058901\\
45600	0.608223483126494\\
46300	0.609999903128482\\
47300	0.593145885155536\\
50300	0.512239145115018\\
51100	0.49502722696343\\
52200	0.509727267941344\\
53100	0.505547444307012\\
54600	0.493203901452944\\
57200	0.505058683324023\\
58500	0.520826097999816\\
59400	0.557205687568057\\
61800	0.616251654035295\\
62500	0.626620377210202\\
64200	0.611336011774256\\
65700	0.674696481990395\\
66500	0.681599730538437\\
68100	0.681868980347645\\
69500	0.671708782305359\\
70500	0.673127665431821\\
72000	0.727377677467302\\
72600	0.722506904785405\\
74200	0.696458573074779\\
75000	0.719600608106703\\
75600	0.72437704836193\\
77000	0.719115437983419\\
79300	0.682844770592055\\
81200	0.689576797740301\\
81700	0.697200025635539\\
82400	0.708772279787809\\
83000	0.697725121644908\\
83900	0.672731394210132\\
84400	0.688830460203462\\
85000	0.713301111376495\\
85500	0.724880469308118\\
86300	0.757165230417741\\
87000	0.754076830708073\\
87600	0.733397638803581\\
88200	0.71234345297853\\
89600	0.677782735903747\\
90300	0.702774516132195\\
90900	0.716221748865792\\
91700	0.709322302936926\\
92500	0.705918633029796\\
93500	0.712971182583715\\
95000	0.689005473206635\\
96500	0.721774116696906\\
96600	0.720331210322911\\
};
\addlegendentry{Loopback}

\addplot [color=mycolor3, line width=1.5pt]
  table[row sep=crcr]{%
0	0.393025281984592\\
1000	0.40307692476199\\
2700	0.359219635356567\\
3500	0.311912032455439\\
4500	0.232195744494675\\
6500	0.0712322296167258\\
7500	0.0314137057284825\\
8100	0.0195459882816067\\
8500	0.0268541776167694\\
9400	0.0740038813528372\\
11400	0.169819667920819\\
12100	0.182689556124387\\
13200	0.176201576090534\\
14100	0.153893114809762\\
15400	0.106616851015133\\
16400	0.0471579089207808\\
17200	0\\
22300	0.00721122795948759\\
22700	0.0450162592460401\\
23100	0.0596591965440894\\
23600	0.0550702804030152\\
24200	0.0293845167907421\\
24800	0\\
25300	0.00508889046614058\\
25800	0.0325402408780064\\
26300	0.0354957071976969\\
26900	0.0174960428557824\\
27500	0\\
28100	0.00373055807722267\\
28500	0.021936650009593\\
29000	0.024722174872295\\
29600	0.0111130388686433\\
30100	0\\
36900	0.00713799569348339\\
37400	0.0211885190801695\\
38300	0.0202816344681196\\
38600	0.0217876068491023\\
39600	0.0786247747164452\\
40200	0.083756130974507\\
41200	0.0695549083320657\\
42700	0.0236572182184318\\
42900	0.0211567923252005\\
43600	0.0385607708012685\\
44400	0.0297458917630138\\
44900	0.0275729002605658\\
45300	0.062112095853081\\
45800	0.093578700456419\\
46500	0.115743000278599\\
47400	0.124891522427788\\
49100	0.11956476890191\\
51100	0.0962929938104935\\
51700	0.11784680832352\\
52300	0.136520701402333\\
53100	0.139771501912037\\
54600	0.137546434096294\\
55500	0.13144887323142\\
55900	0.130961256567389\\
56500	0.138305766668054\\
57300	0.128668764664326\\
58100	0.0965342490671901\\
58400	0.087416767666582\\
59600	0.118179839671939\\
60200	0.103062699345173\\
61300	0.054877364585991\\
62500	0\\
96600	0\\
};
\addlegendentry{Defer}

\end{axis}
\end{tikzpicture}%
\caption{Arrival rate=0.6}
\label{fig:dsgd_prob_0p6}
\end{subfigure}
%\hspace*{\fill}
%\hspace{.05\textwidth}

\begin{subfigure}{0.45\textwidth}
\vspace{20pt}
% This file was created by matlab2tikz.
%
%The latest updates can be retrieved from
%  http://www.mathworks.com/matlabcentral/fileexchange/22022-matlab2tikz-matlab2tikz
%where you can also make suggestions and rate matlab2tikz.
%
\definecolor{mycolor1}{rgb}{0.00000,0.44700,0.74100}%
\definecolor{mycolor2}{rgb}{0.85000,0.32500,0.09800}%
\definecolor{mycolor3}{rgb}{0.92900,0.69400,0.12500}%
\begin{tikzpicture}

\begin{axis}[%
width=2.7in,
height=1.2in,
at={(0in,0in)},
scale only axis,
xmin=0,
xmax=100000,
xlabel={Simulation Time},
ymin=0,
ymax=1,
ylabel={Porbability Vector},
axis background/.style={fill=white},
xmajorgrids,
ymajorgrids,
legend style={nodes={scale=0.7, transform shape}, legend cell align=left, align=left, draw=white!15!black}
]
\addplot [color=mycolor1, dashdotted, line width=1.5pt]
  table[row sep=crcr]{%
0	0.431246327396366\\
1300	0.477540239138762\\
2300	0.532306279885233\\
3600	0.630598970383289\\
4500	0.714975958850118\\
4800	0.740611018118216\\
6500	0.818150180552038\\
7500	0.880388880003011\\
8400	0.951926320398343\\
8900	1\\
12000	0.993234425812261\\
13100	0.950647597564966\\
13900	0.896201867100899\\
15100	0.800410954077961\\
15800	0.725353707588511\\
16700	0.608168248829315\\
18200	0.413907699723495\\
19500	0.262182581893285\\
21700	0.0183776769699762\\
21900	0\\
97100	0.00850077433278784\\
97800	0.0157849267561687\\
};
\addlegendentry{Retransmit}

\addplot [color=mycolor2, dotted, line width=1.5pt]
  table[row sep=crcr]{%
0	0.251638111541979\\
1400	0.208830504430807\\
2500	0.155040372628719\\
3900	0.0648415928735631\\
4800	0\\
21500	0.00715207602479495\\
22200	0.0332792150293244\\
23200	0.0947459056478692\\
23800	0.142039515820215\\
25500	0.29814486973919\\
26800	0.394401968384045\\
27400	0.416203709013644\\
29300	0.445517755695619\\
30100	0.43210988593637\\
31300	0.390065942236106\\
32100	0.346205899113556\\
33800	0.214912073657615\\
34800	0.109324426928652\\
35400	0.0334893987310352\\
35700	0\\
97800	0\\
};
\addlegendentry{Loopback}

\addplot [color=mycolor3, line width=1.5pt]
  table[row sep=crcr]{%
0	0.317115561076207\\
2500	0.298599122732412\\
4800	0.259388981881784\\
6500	0.181849819447962\\
7500	0.119611119996989\\
8400	0.0480736796016572\\
8900	0\\
12000	0.00676557418773882\\
13100	0.0493524024350336\\
13900	0.103798132899101\\
15100	0.199589045922039\\
15800	0.274646292411489\\
16700	0.391831751170685\\
18200	0.586092300276505\\
19500	0.737817418106715\\
21100	0.912905192759354\\
21800	0.977213777980069\\
22000	0.976304467607406\\
22800	0.930781351416954\\
23800	0.857960484179785\\
25500	0.70185513026081\\
26800	0.605598031615955\\
27400	0.583796290986356\\
29300	0.554482244304381\\
30100	0.56789011406363\\
31300	0.609934057763894\\
32100	0.653794100886444\\
33800	0.785087926342385\\
34800	0.890675573071348\\
35400	0.966510601268965\\
35700	1\\
97100	0.991499225667212\\
97800	0.984215073243831\\
};
\addlegendentry{Defer}

\end{axis}
\end{tikzpicture}%
\caption{Arrival rate=3.0}
\label{fig:dsgd_prob_3p0}
\end{subfigure}

\vspace{20pt}
\caption{Probability convergence for $L=10$, $\overline{P}=7$, Zipf Popularity (Zipf exponent =1), Rayleigh fading with mean, 0.1 and 1.0 for bad and good users respectively.}
\end{figure}
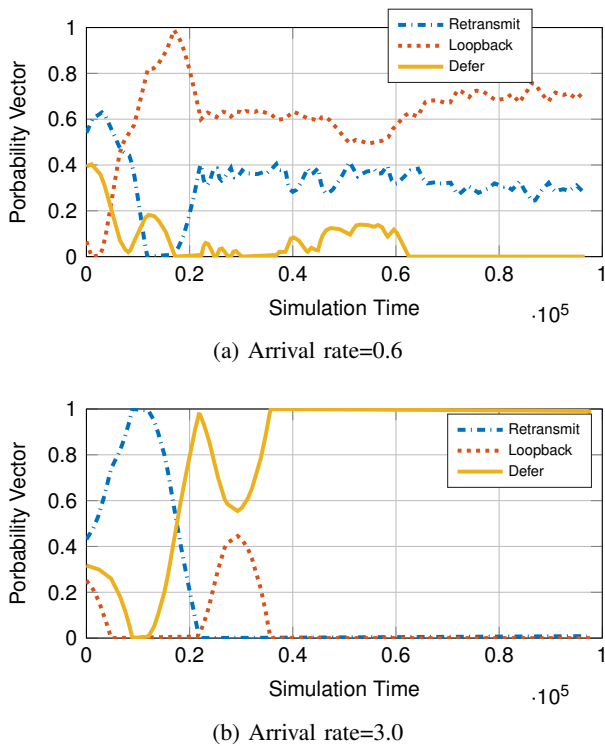

We see in Figure \ref{fig:dsgd_cp} that different queueing strategies are optimal at different rates for a constant transmit power $\overline{P}=7$ under fading. This is the typical case in practical systems. Depending on the request load the system might need to adapt the queueing and service strategy. DSGD does precisely this. We can see in Figure \ref{fig:dsgd_conv} that the algorithm converges to the optimal mean sojourn time for the given power policy. We use constant transmit power policy. Epochs $0$ to $10^4$ are the initial training phase and the algorithm starts learning thereafter and eventually converges. The policy chosen by the algorithm for arrival rates $0.6$ and $3.0$  are given in Figures \ref{fig:dsgd_prob_0p6} and \ref{fig:dsgd_prob_3p0}, respectively. \CE We see that for rate $3.0$, the algorithm converges to the defer strategy since it has the lowest mean sojourn time for this rate (Fig \ref{fig:dsgd_cp})\CE. For rate $0.6$ however we see that DSGD gives a mixed policy with positive probabilities to retransmit and loopback and zero probability to defer. This is because both retransmit and loopback have the same mean delay performance and the defer strategy performs poorly. This is the case where more than one optimal solution may be available and the algorithm may converge to one or oscillate between different optimal points, as neural network training progresses. \CE The simulations show that the DSGD algorithm chooses the best among the three queueing policies or an equivalent mixed policy for different system statistics (arrival rates). This shows that the DSGD adapts to the system statistics which is very important in a practical system. We will see in subsequent sections that the adaptability of DSGD is very useful in cross-layer system optimization of the Multicast network.  \CE

%\begin{figure}[h!]
%\centering
%%\includegraphics[trim={.8cm 10cm 1cm 10cm},clip,height=5cm,width=8.5cm]{JournalFigures/DSGD_Conv.pdf}
%\input{JournalFigures/DSGD_Conv.tex}
%\caption{DSGD Convergence of Mean Sojourn Time with the Optimal Policy. $L=10$, $\overline{P}=7$, Zipf Popularity (Zipf exponent =1), Rayleigh fading with mean, 0.1 and 1.0 for bad and good users respectively.}
%\label{fig:dsgd_conv}
%\end{figure}

\subsection{\CE Optimal power control (AC-DQN vs MADS):}
  \CE We use the system setting of small user case, specified in \ref{sec:system_4}, since running MADS for higher number of users is computationally prohibitive. We use uniform popularity profile for the file requests. We also use uniform distribution for fading. This is just for the convenience of calculations of state probabilities, $\{q_k\}$,  in MADS as done in \cite{wcnc}. We compare the performance of AC-DQN and MADS for this system. We demonstrate our algorithm with more realistic distribution in the next section. 
  
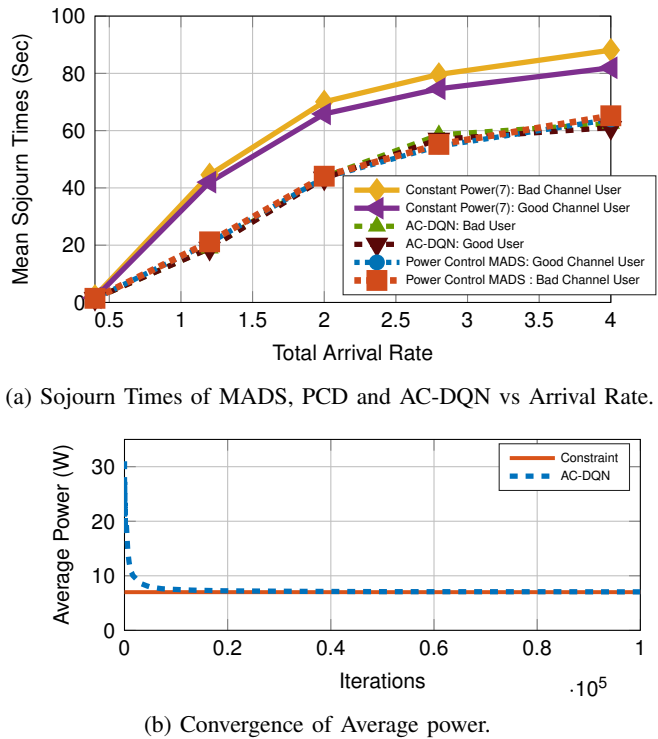
\begin{figure}[t]
\centering
\begin{subfigure}{0.5\textwidth}
\centering
% This file was created by matlab2tikz.
%
%The latest updates can be retrieved from
%  http://www.mathworks.com/matlabcentral/fileexchange/22022-matlab2tikz-matlab2tikz
%where you can also make suggestions and rate matlab2tikz.
%
\definecolor{mycolor1}{rgb}{0.90980,0.67451,0.12549}%
\definecolor{mycolor2}{rgb}{0.92941,0.69412,0.12549}%
\definecolor{mycolor3}{rgb}{0.49412,0.18431,0.55686}%
\definecolor{mycolor4}{rgb}{0.34902,0.03137,0.03137}%
\definecolor{mycolor5}{rgb}{0.00000,0.44706,0.70980}%
\definecolor{mycolor6}{rgb}{0.81961,0.29412,0.10588}%
\begin{tikzpicture}

\begin{axis}[%
width=2.7in,
height=1.5in,
at={(0in,0in)},
scale only axis,
xmin=0.4,
xmax=4,
xlabel={Total Arrival Rate},
ymin=0,
ymax=100,
ylabel={Mean Sojourn Times (Sec)},
axis background/.style={fill=white},
xmajorgrids,
ymajorgrids,
legend style={nodes={scale=0.6, transform shape}, at={(0.48,0.025)}, anchor=south west, legend cell align=left, align=left, draw=white!15!black}
]
\addplot [color=mycolor1, line width=2.0pt, mark size=3.0pt, mark=diamond*, mark options={solid, fill=mycolor2, mycolor1}]
  table[row sep=crcr]{%
0.400000000000006	2.167\\
1.2	44.577\\
2	70.056\\
2.8	79.628\\
4	88.104\\
};
\addlegendentry{Constant Power(7): Bad Channel User}

\addplot [color=mycolor3, line width=2.0pt, mark size=3.0pt, mark=triangle*, mark options={solid,rotate=90, fill=mycolor3, mycolor3}]
  table[row sep=crcr]{%
0.400000000000006	1.60899999999999\\
1.2	42.014\\
2	65.842\\
2.8	74.624\\
4	81.954\\
};
\addlegendentry{Constant Power(7): Good Channel User}

\addplot [color=red!40!green, dashed, line width=2.0pt, mark=triangle*, mark options={solid, fill=red!40!green, red!40!green}]
  table[row sep=crcr]{%
0.399999999999999	1.408\\
1.2	18.94\\
2	43.825\\
2.8	58.513\\
4	62.487\\
};
\addlegendentry{AC-DQN: Bad User}

\addplot [color=mycolor4, dashed, line width=2.0pt, mark size=3.0pt, mark=triangle*, mark options={solid, rotate=180, fill=mycolor4, mycolor4}]
  table[row sep=crcr]{%
0.399999999999999	1.37\\
1.2	18.861\\
2	43.375\\
2.8	56.969\\
4	61.059\\
};
\addlegendentry{AC-DQN: Good User}

\addplot [color=mycolor5, dotted, line width=2.0pt, mark=*, mark options={solid, fill=mycolor5, mycolor5}]
  table[row sep=crcr]{%
0.399999999999999	1.373\\
1.2	20.826\\
2	43.558\\
2.8	54.554\\
4	63.939\\
};
\addlegendentry{Power Control MADS: Good Channel User}

\addplot [color=mycolor6, dotted, line width=2.5pt, mark size=2.7pt, mark=square*, mark options={solid, fill=mycolor6, mycolor6}]
  table[row sep=crcr]{%
0.400000000000006	1.446\\
1.2	21.161\\
2	44.085\\
2.8	55.19\\
4	65.186\\
};
\addlegendentry{Power Control MADS : Bad Channel User}

\end{axis}
\end{tikzpicture}%
\caption{Sojourn Times of MADS, PCD and AC-DQN vs Arrival Rate.\\}
\label{fig:mads_comp_del}
\end{subfigure}

\begin{subfigure}{0.4\textwidth}
\vspace{20pt}
\centering
% This file was created by matlab2tikz.
%
%The latest updates can be retrieved from
%  http://www.mathworks.com/matlabcentral/fileexchange/22022-matlab2tikz-matlab2tikz
%where you can also make suggestions and rate matlab2tikz.
%
\definecolor{mycolor1}{rgb}{0.85000,0.32500,0.09800}%
\definecolor{mycolor2}{rgb}{0.00000,0.44706,0.74118}%
\begin{tikzpicture}

\begin{axis}[%
width=2.7in,
height=1.0in,
at={(0in,0in)},
scale only axis,
xmin=0,
xmax=100000,
xlabel={Iterations},
ymin=0,
ymax=35,
ylabel={Average Power (W)},
axis background/.style={fill=white},
xmajorgrids,
ymajorgrids,
legend style={nodes={scale=0.6, transform shape}, legend cell align=left, align=left, draw=white!15!black}
]
\addplot [color=mycolor1, line width=1.5pt]
  table[row sep=crcr]{%
1	7\\
100000	7\\
};
\addlegendentry{Constraint}

\addplot [color=mycolor2, dashed, line width=2.0pt]
  table[row sep=crcr]{%
1	31\\
2	29.125\\
3	20.5833333333285\\
4	24.1875\\
5	22.8500000000058\\
6	27.125\\
7	27.1428571428551\\
8	26.3125\\
9	23.527777777781\\
10	26.0249999999942\\
11	27.2272727272793\\
13	23.6153846153902\\
14	23.3928571428551\\
16	21.125\\
17	22.7352941176505\\
18	21.819444444438\\
19	22.9736842105194\\
23	22.1521739130403\\
24	23.25\\
25	22.9100000000035\\
26	23.8942307692341\\
27	23.1388888888905\\
28	23.4196428571449\\
29	22.646551724145\\
30	22.4750000000058\\
31	22.75\\
33	21.4545454545441\\
34	22.1102941176505\\
36	21.3888888888905\\
37	21.0270270270266\\
39	22.1987179487187\\
42	21.2857142857101\\
43	21.4244186046562\\
45	21.1055555555504\\
52	20.2932692307659\\
54	19.9861111111095\\
55	20.4181818181823\\
58	19.4956896551739\\
59	19.9872881355986\\
61	19.3934426229534\\
64	19.76953125\\
66	19.5454545454559\\
69	19.3369565217436\\
72	19.315972222219\\
73	19.7157534246508\\
75	19.736666666664\\
83	20.5662650602462\\
84	20.3363095238165\\
86	20.5959302325646\\
88	20.46875\\
90	20.6222222222277\\
95	19.8894736842049\\
98	20.0739795918344\\
105	19.7619047619082\\
108	20.391203703708\\
110	20.434090909097\\
114	20.2850877192977\\
116	20.3943965517246\\
125	20.5420000000013\\
131	20.0648854961764\\
133	20.1635338345804\\
139	19.7589928057569\\
142	19.6285211267677\\
146	19.5445205479482\\
150	19.8333333333285\\
158	19.3227848101233\\
162	19.043209876545\\
172	18.672965116275\\
176	19.0923295454559\\
182	19.0164835164906\\
193	19.1541450777149\\
198	19.1755050505017\\
205	19.2280487804819\\
208	19.25\\
210	19.3023809523875\\
215	19.4941860465187\\
223	19.2746636771335\\
348	19.5926724137971\\
367	19.2568119891075\\
624	14.7596153846098\\
667	14.359070464765\\
778	13.4145244215906\\
832	12.9353966346098\\
891	12.487654320983\\
1105	11.3938914027094\\
1347	10.6648106904177\\
1364	10.6367302052822\\
1448	10.4442334254127\\
1471	10.3774643099896\\
1629	10.0417434008559\\
3042	8.67504930966243\\
3276	8.54517704517639\\
4486	8.09629959875019\\
4530	8.09359823400155\\
5103	7.91710758378031\\
7204	7.63561910050339\\
7770	7.59424066923384\\
8339	7.57111164407979\\
8973	7.53315502061741\\
12884	7.37674635206349\\
13025	7.37136276392266\\
19189	7.24507530356641\\
19577	7.24521121724683\\
20939	7.23213859305542\\
21168	7.22778486394964\\
31327	7.16330960513733\\
33730	7.1487918766652\\
51191	7.0879256119224\\
51836	7.08874141523847\\
77655	7.07577425792988\\
83792	7.07994199923996\\
93759	7.05882901907898\\
94673	7.05817392499011\\
99980	7.05484346869343\\
};
\addlegendentry{AC-DQN}

\end{axis}
\end{tikzpicture}%
\caption{Convergence of Average power.}	
\label{fig:AC-DQN_power_4}
\end{subfigure}
\vspace{20pt}
\caption{AC-DQN Performance in 1-LB system with $L=4$, $\overline{P}=7$, Uniform Popularity, Uniform fading.}
\end{figure}

  We use the Loopback queueing strategy for demonstrating AC-DQN. We will see in subsequent sections that AC-DQN works even for other queueing strategies\CE. We split the users in two equal sized groups, one group has good channel statistics and the other bad channel statistics, to show the advantage of power control. We compare both power control policies with a constant power control policy, where the transmit power $P_t$ is fixed to $P_t=\overline{P}$, to indicate the gain due to power control. Figure \ref{fig:mads_comp_del} shows a comparison of mean sojourn times of Constant Power Policy, $P_t=\overline{P}$, MADS and AC-DQN. Further, Figure \ref{fig:AC-DQN_power_4} shows convergence of average power to $\overline{P}$ for AC-DQN. We see from Figure \ref{fig:AC-DQN_power_4} that AC-DQN \CE achieves the same mean sojourn time \CE as that by MADS, while maintaining the average power constraint. \CE Further we demonstrate power control by AC-DQN for a scaled up system with 20 users.\CE

%
%\begin{figure}[h]
%\centering
%%\includegraphics[trim={1cm 7.5cm 1cm 8cm},clip,height=5cm,width=8.5cm]{DeepRL_Figs/MADS_PCD_APCD_Delay_4.pdf}
%\input{DeepRL_Figs/MADS_PCD_APCD_Delay_4.tex}
%\caption{Sojourn Times of MADS, PCD and AC-DQN vs Arrival Rate for Loopback Queue. $L=4$, $\overline{P}=7$, Uniform Popularity, Uniform fading.}
%\label{fig:mads_comp_del}
%\end{figure}
%
%\begin{figure}[!h]
%\centering
%%	\includegraphics[trim={1cm 13.3cm 1.2cm 8.5cm},clip,height=3cm,width=8.5cm]{DeepRL_Figs/PC_4.pdf}
%\input{DeepRL_Figs/PC_4.tex}
%\caption{Convergence of Average power with Iteration for AC-DQN, $L=4,\overline{P}=7$.}	
%\label{fig:AC-DQN_power_4}
%\end{figure}

\subsection{AC-DQN performance in a Scaled Network:}
To show the scalability of AC-DQN. We simulate  the relatively complex system mentioned in large user case, section \ref{sec:system_20}. We use Zipf Popularity, Rayleigh fading (refer Section \ref{sec:system_20}) to analyse AC-DQN on a more realistic system. We use Loopback queueing strategy at the server and run the simulation for the average power constraint $\overline{P}=7$. We see in Figure \ref{fig:AC-DQN_comp_del} that the AC-DQN gives, drastic improvement (around ~50 percent) over constant power case. AC-DQN achieves this while maintaining the average power, by learning the Lagrange constant as seen in Figure \ref{fig:Lagrange_20}. Figure \ref{fig:APCD_power_20} shows the convergence of average power of AC-DQN to the average power constraint, $\overline{P}$, for arrival rate of 1.0 requests per sec in the same simulation run.
\begin{figure}
\noindent\begin{subfigure}{0.45\textwidth}
\centering
% This file was created by matlab2tikz.
%
%The latest updates can be retrieved from
%  http://www.mathworks.com/matlabcentral/fileexchange/22022-matlab2tikz-matlab2tikz
%where you can also make suggestions and rate matlab2tikz.
%
\definecolor{mycolor1}{rgb}{0.91765,0.68235,0.12549}%
\definecolor{mycolor2}{rgb}{0.49020,0.18039,0.55686}%
\definecolor{mycolor3}{rgb}{0.49412,0.18431,0.55686}%
\definecolor{mycolor4}{rgb}{0.35294,0.03529,0.03529}%
\definecolor{mycolor5}{rgb}{0.30980,0.02745,0.02745}%
\begin{tikzpicture}

\begin{axis}[%
width=2.7in,
height=1.5in,
at={(0in,0in)},
scale only axis,
xmin=0.2,
xmax=1,
xlabel={Total Arrival Rate},
ymin=0,
ymax=153.696617324234,
ylabel={Mean Sojourn Times(sec)},
axis background/.style={fill=white},
xmajorgrids,
ymajorgrids,
legend style={nodes={scale=0.6, transform shape}, at={(0.024,0.7)}, anchor=south west, legend cell align=left, align=left, draw=white!15!black}
]
\addplot [color=mycolor1, line width=2.0pt, mark size=3pt, mark=diamond*, mark options={solid, fill=mycolor1, mycolor1}]
  table[row sep=crcr]{%
0.199999999999989	7.13337602415851\\
0.400000000000006	21.1346457070742\\
0.599999999999994	67.2742461733176\\
0.800000000000011	115.034473439055\\
1	153.696617324234\\
};
\addlegendentry{Constant Power(7) : Bad Users}

\addplot [color=mycolor2, line width=2.0pt, mark size=2.7pt, mark=square*, mark options={solid, fill=mycolor3, mycolor2}]
  table[row sep=crcr]{%
0.200000000000003	2.49635418626117\\
0.400000000000006	9.58906493820993\\
0.599999999999994	31.4912665875862\\
0.799999999999997	51.9268815792685\\
1	68.1371716816205\\
};
\addlegendentry{Constant Power(7) : Good Users}

\addplot [color=red!40!green, dashed, line width=2.5pt, mark size=2.7pt, mark=triangle*, mark options={solid, fill=red!40!green, red!40!green}]
  table[row sep=crcr]{%
0.200000000000003	3.866\\
0.400000000000006	7.077\\
0.599999999999994	15.813\\
0.799999999999997	40.848\\
1	65.629\\
};
\addlegendentry{AC-DQN: Bad Users}

\addplot [color=mycolor4, dashed, line width=2.5pt, mark size=2.7pt, mark=triangle*, mark options={solid, rotate=180, fill=mycolor5, mycolor4}]
  table[row sep=crcr]{%
0.200000000000003	1.862\\
0.399999999999999	3.897\\
0.600000000000001	10.319\\
0.799999999999997	27.075\\
1	42.341\\
};
\addlegendentry{AC-DQN : Good Users}

\end{axis}
\end{tikzpicture}%
\caption{Constant Power and AC-DQN vs Arrival Rate}
\label{fig:AC-DQN_comp_del}
\end{subfigure}
%\hspace*{\fill}

\begin{subfigure}{0.45\textwidth}
\vspace{20pt}
\centering
% This file was created by matlab2tikz.
%
%The latest updates can be retrieved from
%  http://www.mathworks.com/matlabcentral/fileexchange/22022-matlab2tikz-matlab2tikz
%where you can also make suggestions and rate matlab2tikz.
%
\definecolor{mycolor1}{rgb}{0.00000,0.44700,0.74100}%
\definecolor{mycolor2}{rgb}{0.85000,0.32500,0.09800}%
\begin{tikzpicture}

\begin{axis}[%
width=2.7in,
height=1.0in,
at={(0in,0in)},
scale only axis,
xmin=0,
xmax=100000,
xlabel={Iterations},
ymin=0,
ymax=21.0833333333333,
ylabel={Average Power},
axis background/.style={fill=white},
xmajorgrids,
ymajorgrids,
legend style={nodes={scale=0.6, transform shape}, legend cell align=left, align=left, draw=white!15!black}
]
\addplot [color=mycolor1, dashed, line width=1.5pt]
  table[row sep=crcr]{%
1	12.25\\
2	18\\
3	21.0833333333285\\
4	20.1875\\
7	15.6071428571449\\
8	19.125\\
10	16.7749999999942\\
12	14.2291666666715\\
13	13.4038461538439\\
15	12.4666666666599\\
16	12.96875\\
17	12.5735294117621\\
18	13.819444444438\\
19	13.1842105263204\\
20	13.8874999999971\\
22	12.8522727272793\\
23	13.6413043478242\\
24	14.8958333333285\\
26	15.1057692307659\\
27	16.3425925925985\\
28	16.4910714285652\\
29	16.0862068965507\\
31	16.8064516129089\\
32	16.6640625\\
33	16.1969696969754\\
34	16.2352941176505\\
35	15.8428571428522\\
36	16.159722222219\\
37	16.1959459459467\\
38	15.861842105267\\
39	16.1538461538439\\
42	15.2916666666715\\
46	14.6684782608645\\
47	15.1914893617068\\
49	15\\
50	14.7350000000006\\
52	14.7884615384683\\
54	15.6157407407445\\
55	15.6499999999942\\
57	15.1491228070226\\
59	15.2076271186379\\
60	14.9833333333372\\
61	15.0737704917992\\
62	14.8870967741968\\
64	15.33203125\\
66	14.9015151515196\\
68	15.2647058823495\\
69	15.0615942028962\\
70	15.539285714287\\
78	14.9903846153902\\
84	15.7648809523816\\
101	15.3712871287134\\
102	15.2671568627411\\
103	15.5898058252496\\
107	15.2289719626133\\
108	15.5370370370365\\
117	15.2521367521404\\
119	15.2773109243717\\
120	15.5541666666686\\
134	15.3936567164201\\
136	15.498161764699\\
142	15.4859154929582\\
144	15.399305555562\\
145	15.5948275862029\\
150	15.4799999999959\\
152	15.6924342105194\\
165	15.495454545453\\
167	15.6047904191655\\
200	15.6662499999948\\
205	15.9317073170678\\
215	16.1488372093008\\
217	16.2695852534525\\
225	16.1100000000006\\
229	16.2096069869003\\
238	15.9978991596581\\
240	16.1802083333314\\
247	16.0455465587002\\
250	16.1729999999952\\
254	16.2982283464517\\
260	16.3673076923005\\
263	16.3925855513371\\
268	16.3563432835799\\
272	16.3391544117621\\
288	16.175347222219\\
298	15.9723154362437\\
317	15.9179810725618\\
322	16.1234472049691\\
344	15.9767441860458\\
349	15.9484240687743\\
361	15.9958448753459\\
367	16.0613079019095\\
460	15.9630434782594\\
483	16.3079710144957\\
489	16.3614519427356\\
499	16.6317635270534\\
504	16.6736111111095\\
508	16.7263779527566\\
516	16.7199612403056\\
533	16.7462476547807\\
542	16.6891143911489\\
558	16.6608422939025\\
574	16.4812717769964\\
578	16.4848615916999\\
620	15.7463709677395\\
628	15.6731687898136\\
660	15.0435606060637\\
708	14.3562853107287\\
714	14.2948179271771\\
844	12.5749407582916\\
899	12.1123470522871\\
932	11.8825107296143\\
990	11.5290404040425\\
998	11.5082665330701\\
1045	11.2492822966451\\
1083	11.0436288088677\\
1134	10.8097442680737\\
1758	9.50625711034809\\
1771	9.49717673630221\\
1841	9.40793047257466\\
1863	9.42257112184598\\
1931	9.34632314862392\\
1944	9.3533950617275\\
2189	9.01359068068268\\
2281	8.90881192460074\\
2433	8.87104397862277\\
2456	8.84527687296213\\
2731	8.62046869278129\\
2748	8.63982896652305\\
2883	8.5188172042981\\
2963	8.52657779277069\\
3120	8.36474358974374\\
3151	8.34711202792823\\
3318	8.33092224231223\\
3393	8.29634541703854\\
3559	8.19190783928207\\
3612	8.18016334441199\\
3679	8.1678445229627\\
3820	8.13272251309536\\
3923	8.1421106296184\\
4110	8.08053527980519\\
4326	8.05796347664727\\
4573	7.93888038487057\\
4654	7.92125053716882\\
4855	7.89850669412408\\
4937	7.86996151508356\\
5156	7.84193173002859\\
5313	7.81032373424387\\
5508	7.80224219316733\\
5771	7.77174666435167\\
6017	7.7262755526026\\
6133	7.72974074678496\\
6394	7.68712855802733\\
6476	7.69603150092007\\
6750	7.65544444444822\\
7069	7.62933229593909\\
7331	7.62539217023004\\
7488	7.59331597221899\\
7790	7.56749037226837\\
7843	7.56990309830871\\
8232	7.52335398444848\\
8284	7.54065065186296\\
8628	7.49235048679111\\
10973	7.38672195388062\\
11418	7.36862848134479\\
11514	7.35854177523288\\
11985	7.3521068001719\\
15443	7.27706727967598\\
15573	7.26685609709239\\
16074	7.26391999502084\\
16930	7.25404607206292\\
21824	7.20660740468884\\
22018	7.21003270051733\\
23067	7.16956041097001\\
24985	7.16108665199135\\
25994	7.15563206894149\\
26222	7.15660514072806\\
27285	7.1505589151493\\
28541	7.14144563960144\\
29667	7.14296187683067\\
30197	7.1380103983829\\
31401	7.13653227603936\\
31773	7.13115695716988\\
32354	7.13198522593302\\
40011	7.09356801879767\\
41696	7.07988775901322\\
41936	7.07482830980734\\
44298	7.06090568423679\\
46400	7.06337284483016\\
48289	7.06035536043055\\
51092	7.03891509433743\\
51534	7.03727636123949\\
53635	7.03499580497737\\
57442	7.027636572544\\
58029	7.02885195333511\\
60741	7.02918539372331\\
63327	7.02697901369538\\
85805	6.99474972320604\\
86536	6.98956214754435\\
99985	6.98518527779379\\
};
\addlegendentry{AC-DQN}

\addplot [color=mycolor2, line width=1.5pt]
  table[row sep=crcr]{%
1	7\\
100000	7\\
};
\addlegendentry{Constraint}

\end{axis}
\end{tikzpicture}%	
\caption{Average Power Convergence.}
\label{fig:APCD_power_20}
\end{subfigure}
\hspace{.025\textwidth}
%\hspace*{\fill}

\begin{subfigure}{0.45\textwidth}
\vspace{20pt}
\centering
% This file was created by matlab2tikz.
%
%The latest updates can be retrieved from
%  http://www.mathworks.com/matlabcentral/fileexchange/22022-matlab2tikz-matlab2tikz
%where you can also make suggestions and rate matlab2tikz.
%
\definecolor{mycolor1}{rgb}{0.00000,0.44700,0.74100}%
\begin{tikzpicture}

\begin{axis}[%
width=2.7in,
height=1.0in,
at={(0in,0in)},
scale only axis,
xmin=0,
xmax=100000,
xlabel={Iterations},
ymin=0,
ymax=0.2,
ytick={0,0.1,0.2},
ylabel={$\text{Lagrange(}\beta\text{)}$},
axis background/.style={fill=white},
xmajorgrids,
ymajorgrids,
legend style={nodes={scale=0.6, transform shape}, legend cell align=left, align=left, draw=white!15!black}
]
\addplot [color=mycolor1, line width=1.5pt]
  table[row sep=crcr]{%
1	0.0216948066954501\\
105	0.131058387545636\\
134	0.155308584537124\\
161	0.173504011792829\\
180	0.182254459010437\\
203	0.189005383959739\\
229	0.191345540602924\\
254	0.189343224701588\\
287	0.181224234620458\\
479	0.122957009094534\\
530	0.115116870802012\\
728	0.0942375276499661\\
865	0.0887848529673647\\
996	0.0922660417709267\\
1168	0.103739883561502\\
1363	0.102717012006906\\
1418	0.104098787647672\\
1478	0.106889013200998\\
1555	0.107995950849727\\
1626	0.104344537670841\\
1771	0.097307325937436\\
1876	0.0945758370653493\\
1917	0.0950292465422535\\
2033	0.104775269486709\\
2102	0.111230669906945\\
2168	0.111614440247649\\
2380	0.105472187919077\\
2427	0.104411825173884\\
2517	0.104495234278147\\
2694	0.0933767110109329\\
2761	0.0910016352863749\\
2853	0.0953004107432207\\
2943	0.100447692166199\\
3020	0.0977869764756178\\
3153	0.0911647908797022\\
3358	0.0918701261398382\\
3483	0.101282178628026\\
3535	0.104874386495794\\
3645	0.104571931107785\\
3820	0.107291512773372\\
3902	0.108713148802053\\
3965	0.105815212460584\\
4109	0.0936110361508327\\
4193	0.0915927918103989\\
4293	0.0966900809580693\\
4409	0.0991422512306599\\
4470	0.0951076974452008\\
4539	0.0923215463990346\\
4606	0.0912674187420635\\
4832	0.0913853923120769\\
4922	0.0891782290418632\\
4985	0.089727007158217\\
5042	0.0929705465823645\\
5221	0.105884395044995\\
5363	0.102247868300765\\
5450	0.10038096975768\\
5600	0.100687962709344\\
6122	0.103051444646553\\
6335	0.101095485471888\\
6459	0.104134800814791\\
6575	0.105746533867205\\
6640	0.105794936651364\\
6840	0.108139218660654\\
6913	0.108561426808592\\
7003	0.105911640173872\\
7211	0.096339905532659\\
7306	0.0986578398587881\\
7463	0.104822432476794\\
7560	0.102420302195242\\
7748	0.100187831907533\\
8060	0.0975803665060084\\
8287	0.0950933541171253\\
8480	0.0989284111710731\\
8559	0.0990065895312\\
8953	0.0859511463786475\\
9051	0.0876056171109667\\
9172	0.0949807862489251\\
9297	0.103548811574001\\
9375	0.105711033043917\\
9502	0.105127195813111\\
9743	0.098944369572564\\
9878	0.089617318008095\\
9967	0.0903089996136259\\
10335	0.101930525474017\\
10484	0.0967722216446418\\
10709	0.0877626321162097\\
11136	0.0863562924350845\\
11280	0.0852355946990428\\
11399	0.089668033979251\\
11481	0.0920734412502497\\
11590	0.0908967290160945\\
11819	0.0857356788619654\\
11910	0.0911556313803885\\
12034	0.100008519366384\\
12177	0.0988230053189909\\
12464	0.0942359786713496\\
12639	0.0973315105657093\\
12776	0.0945379969925852\\
13065	0.0881149807391921\\
13185	0.0905485278199194\\
13311	0.0981472966523143\\
13415	0.101717468627612\\
13560	0.10382389750157\\
13929	0.103504906001035\\
14110	0.0998079492273973\\
14265	0.0997784270293778\\
14566	0.101602539318264\\
14862	0.0931542594917119\\
15274	0.0906100033025723\\
15447	0.0905088946747128\\
15753	0.0939113172498764\\
16021	0.0943894533993443\\
16181	0.0938829114020336\\
16444	0.0917356605059467\\
16609	0.0912913960346486\\
16995	0.0908562291879207\\
17211	0.0935733123769751\\
17377	0.096740868219058\\
17595	0.0975507739349268\\
17836	0.0993532810389297\\
18985	0.0998357641801704\\
19150	0.100514733421733\\
19435	0.0974850362108555\\
19647	0.0983078235149151\\
19845	0.0960100595111726\\
20035	0.0960834700672422\\
20177	0.096045315003721\\
20336	0.0958911614579847\\
20539	0.0944317491084803\\
20951	0.0963717587001156\\
21167	0.098307018604828\\
21318	0.100121481533279\\
21594	0.100417198002106\\
21882	0.0988838765624678\\
22383	0.0907607814733637\\
22628	0.0874801554746227\\
22795	0.0874815013230545\\
23151	0.0833688001439441\\
23618	0.0870083019981394\\
23757	0.0875403475074563\\
23966	0.0884486065770034\\
24371	0.0893134487705538\\
24624	0.0891351877944544\\
24993	0.08705514074245\\
25411	0.0909367208369076\\
25600	0.0894319173967233\\
26734	0.0907637366763083\\
27203	0.0899811131966999\\
27580	0.0876261358644115\\
27798	0.0867269691661932\\
28689	0.0896834154991666\\
29066	0.0900054232915863\\
29442	0.0917169670865405\\
30174	0.0924217040155781\\
30776	0.090115476414212\\
30982	0.0920432775601512\\
31301	0.0916618884220952\\
31752	0.0924805446411483\\
32095	0.0922564609936671\\
32407	0.0923123851098353\\
32939	0.0906226589286234\\
33244	0.0913805836898973\\
33679	0.0929101029469166\\
34659	0.0888715601613512\\
35124	0.0878851539600873\\
36316	0.0870113444689196\\
36641	0.0873300481762271\\
37509	0.0855239246593555\\
38369	0.0892487687087851\\
40410	0.0865511155425338\\
40911	0.0863386301207356\\
42849	0.0828060701896902\\
43799	0.0819379193126224\\
44270	0.0830422677099705\\
44733	0.0832903334521689\\
46002	0.0831658407405484\\
48664	0.0819845974765485\\
51097	0.0798406538378913\\
56347	0.0795838717895094\\
57242	0.0791468972456641\\
71728	0.0786272913537687\\
75347	0.0785319358546985\\
91217	0.0780717433517566\\
99484	0.0780095858644927\\
};
\addlegendentry{$\beta\text{ Convergence}$}

\end{axis}
\end{tikzpicture}%
\caption{Lagrange\\ Convergence.}
\label{fig:Lagrange_20}
\end{subfigure}

\vspace{20pt}
\caption{AC-DQN Performance in 1-LB system with $L=20$, $\overline{P}=7$, Zipf(1) Popularity, Rayleigh fading.}
\end{figure}

%\begin{figure}[h!]
%\centering
%%\includegraphics[trim={.7cm 7.6cm 1.2cm 8.1cm},clip,height=4.5cm,width=8.5cm]{DeepRL_Figs/PCD_APCD_Delay_20.pdf}
%\input{PCD_APCD_Delay_20.tex}
%\caption{Sojourn times for Constant Power and AC-DQN vs Arrival Rate for Loopback Queue. $L=20$, $\overline{P}=7$, Zipf(1) Popularity, Rayleigh fading.}
%\label{fig:AC-DQN_comp_del}
%\end{figure}
%%\begin{figure}[h!]
%%	\includegraphics[trim={0cm 7cm 0cm 7cm},clip,height=5.5cm,width=8.5cm]{DeepRL_Figs/Avg_power_pcd_20.pdf}
%%\label{fig:pcd_power_20}
%%\caption{Convergence of Average Power, PCD \color{red}{This plot will be merged with AC-DQN figure}}		
%%\end{figure}
%\begin{figure}[h!]
%\centering
%%	\includegraphics[trim={1.1cm 13cm 1.4cm 8.5cm},clip,height=3cm,width=8.5cm]{DeepRL_Figs/Avg_power_20.pdf}
%\input{Avg_power_20.tex}	
%\caption{Convergence of Average Power, AC-DQN, $L=20, \overline{P}=7.$}
%\label{fig:APCD_power_20}
%\end{figure}
%\begin{figure}[h!]
%\centering
%%	\includegraphics[trim={.6cm 13.3cm 1.4cm 8.5cm},clip,height=3cm,width=8.5cm]{DeepRL_Figs/Lagrange_20.pdf}	
%\input{Lagrange_20.tex}
%\caption{Convergence of Lagrange multiplier $L=20, \overline{P}=7.$}
%\label{fig:Lagrange_20}
%\end{figure}

\subsection{AC-DQN Tracking Simulations:}
In this section, we show via simulations, the tracking capabilities of AC-DQN for the large user case (section \ref{sec:system_20}). 
%We fix $\eta_1 = 0.001$ and $\eta_2 = 0.00003$. As explained previously, this is important for detecting the change in the environment dynamics faster. In this simulation, we vary the arrival rate over a period of 48 hours. This captures the real world scenario where the request traffic to the base station varies with time of the day. To make the learning harder for our algorithm, we make these rate changes abruptly at every six hours. Specifically we use arrival rates $\lambda= 0.4, 0.8, 0.2, 1.0$ for 1st, 2nd, 3rd and 4th six hour period, respectively. We plot the AC-DQN performance for $\overline{P}=7$ in Figure \ref{fig:track_7}. We calculate the mean sojourn time and average power using a moving average window of size 1000 samples. We observe that for each arrival rate in this simulation, the AC-DQN achieves the corresponding stationary mean sojourn time performance. For instance for $\lambda=0.8$ and $\overline{P}=7$, the values in Figure \ref{fig:AC-DQN_comp_del} and Figure \ref{fig:track_7} are comparable. It is important to note that this performance is achieved while maintaining the average power constraint as can be seen in Figure \ref{fig:track_p_7}. The effect of fixing the learning rates is seen in the small oscillations of average power around $\overline{P}=7$ in Figure \ref{fig:track_p_7}. This is the oscillation in a small neighborhood around the optimal average power. Smaller the step size, lesser the oscillations. 
We demonstrate the importance of constant step sizes for $\eta_1$ and $\eta_2$, and the inability of decaying step sizes to track the changing system statistics. We consider a system where the arrival rates change over a period of 48 hours. We fix $\lambda=1.0$ for first 24 hours. To make the learning harder for our algorithm, we change the rates abruptly, every six hours for next 24 hours as $\lambda=0.6, 0.5, 0.4, 0.8$. This change in time period is just to illustrate the tracking ability in a more emphatic manner. This also captures the real world scenario where the request traffic to the base station varies with time of the day. We fix $\overline{P}=5$. We calculate the mean sojourn time and average power using a moving average window of size 1000 samples. We run the AC-DQN algorithm for this system with: 1) decaying $\eta_1$ and $\eta_2$ satisfying (\ref{eq:two_timeline_learning}) and 2) constant step sizes, $\eta_1=0.001$ and $\eta_2=0.00003$. Rest of the parameters remain same as in the large user case. We see in Figure \ref{fig:decay_const_del} that the AC-DQN with constant step-size almost always outperforms the decaying step size. Specifically, after the first 24 hours the delay reduction is nearly 50 percent for constant step-size. The reason for this is evident from Figures \ref{fig:decay_const_pavg} and  \ref{fig:decay_const_beta}. We see in Figure \ref{fig:decay_const_beta} that the AC-DQN with constant step-size learns the Lagrange constant through out the simulation time, whereas, the AC-DQN decaying step size is unable to learn the  Lagrange constant after the first 24 hours. As can be seen in Figure \ref{fig:decay_const_pavg}, this affects the average power achieved by the AC-DQN with decaying step size. While constant step size maintains the average power constraint of $\overline{P}=5$, the average power achieved by the decaying step-size AC-DQN drops to $4$. Hence, the decaying step-size AC-DQN suffers suboptimal utilization of available power.  Thus in practical systems, only constant step-size AC-DQN will be capable of adapting to the changing system statistics. The effect of fixing the learning rates is seen in the small oscillations of average power around $\overline{P}=5$ in Figure \ref{fig:decay_const_pavg}. This is the oscillation in a small neighborhood around the optimal average power. Smaller the step size, lesser the oscillations.            

%\begin{figure}[h!]
%\centering
%	\includegraphics[trim={.7cm 7.7cm 1.6cm 8cm},clip,height=5.5cm,width=8.5cm]{DeepRL_Figs/track_delay_5.pdf}
%	\caption{AC-DQN Tracking: Delay performance of AC-DQN vs Constant Power Policy, for $L=20$, $\overline{P}=5.$ }	
%\label{fig:track_5}
%\end{figure}
%
%\begin{figure}[h!]
%\centering
%	\includegraphics[trim={1.1cm 13cm 1.6cm 8cm},clip,height=3.5cm,width=8.5cm]{DeepRL_Figs/track_pavg_5.pdf}
%\caption{Tracking of Power Constraint by AC-DQN with tracking for $L=20$, $\overline{P}=5.$}	
%\label{fig:track_p_5}
%\end{figure}

%\begin{figure}[h!]
%\centering
%\includegraphics[trim={.7cm 7.7cm 1.6cm 8cm},clip,height=5.5cm,width=8.5cm]{DeepRL_Figs/track_delay_7.pdf}	
%\caption{AC-DQN Tracking: Delay performance of AC-DQN vs Constant Power Policy, for $L=20$, $\overline{P}=7.$ }
%\label{fig:track_7}
%\end{figure} 
%
%\begin{figure}[h!]
%\centering
%	\includegraphics[trim={1.1cm 13cm 1.6cm 8cm},clip,height=3.5cm,width=8.5cm]{DeepRL_Figs/track_pavg_7.pdf}	
%\caption{Tracking of Power Constraint by AC-DQN with tracking for $L=20$, $\overline{P}=7.$}	
%\label{fig:track_p_7}
%\end{figure}

\begin{figure}
\noindent\begin{subfigure}{0.45\textwidth}
\centering
\input{decay_const_del_2.tex}
\caption{Mean Sojourn Time tracking}
\label{fig:decay_const_del}
\end{subfigure}
%\hspace*{\fill}

\begin{subfigure}{0.45\textwidth}
\vspace{20pt}
\centering
\input{decay_const_pavg.tex}
\caption{Average Power}
\label{fig:decay_const_pavg}
\end{subfigure}

\begin{subfigure}{0.45\textwidth}
\vspace{20pt}
\centering
\input{decay_const_beta.tex}
\caption{Lagrange}
\label{fig:decay_const_beta}
\end{subfigure}

\vspace{20pt}
\caption{AC-DQN Tracking Performance in 1-LB system with with decaying vs constant step-sizes $L=20$, $\overline{P}=7$, Zipf(1) Popularity, Rayleigh fading.}
\end{figure}
%
%\begin{figure}[h!]
%\centering
%%\includegraphics[trim={.7cm 7.6cm 1.2cm 8.1cm},clip,height=4.5cm,width=8.5cm]{DeepRL_Figs/decay_const_del_2.pdf}
%\input{decay_const_del_2.tex}
%\caption{Sojourn times for AC-DQN with decaying vs constant step-sizes for Loopback Queue. $L=20$, $\overline{P}=5$, Zipf(1) Popularity, Rayleigh fading.}
%\label{fig:decay_const_del}
%\end{figure}
%
%\begin{figure}[h!]
%\centering
%%	\includegraphics[trim={1cm 9.5cm 1.2cm 10cm},clip,height=3.5cm,width=8.5cm]{DeepRL_Figs/decay_const_pavg.pdf}	
%\input{decay_const_pavg.tex}
%\caption{Convergence of Average Power for AC-DQN with decaying vs constant step-sizes. $L=20, \overline{P}=5.$}
%\label{fig:decay_const_pavg}
%\end{figure}
%
%\begin{figure}[h!]
%\centering
%%\includegraphics[trim={.4cm 9.6cm 1.2cm 10.1cm},clip,height=3.5cm,width=8.5cm]{DeepRL_Figs/decay_const_beta.pdf}
%\input{decay_const_beta.tex}
%\caption{Convergence of Lagrange for AC-DQN with decaying vs constant step-sizes. $L=20, \overline{P}=5.$}
%\label{fig:decay_const_beta}
%\end{figure}

\subsection{Integrated Optimal Queueing and Power Control using IDA}
\begin{figure}
\centering

\noindent\begin{subfigure}{0.45\textwidth}
\centering
% This file was created by matlab2tikz.
%
%The latest updates can be retrieved from
%  http://www.mathworks.com/matlabcentral/fileexchange/22022-matlab2tikz-matlab2tikz
%where you can also make suggestions and rate matlab2tikz.
%
\definecolor{mycolor1}{rgb}{0.00000,0.44700,0.74100}%
\definecolor{mycolor2}{rgb}{0.85098,0.32941,0.10196}%
\definecolor{mycolor3}{rgb}{1.00000,0.00000,1.00000}%
\begin{tikzpicture}

\begin{axis}[%
width=2.6in,
height=1.5in,
at={(0in,0in)},
scale only axis,
xmin=0,
xmax=3,
xlabel={Total Arrival Rate},
ymin=0,
ymax=180,
ylabel={Mean Sojourn Time (Sec)},
axis background/.style={fill=white},
xmajorgrids,
ymajorgrids,
legend style={nodes={scale=0.7, transform shape}, at={(0.16,0.716)}, anchor=south west, legend cell align=left, align=left, draw=white!15!black}
]
\addplot [color=mycolor1, dashed, line width=2.0pt, mark size=3.0pt, mark=+, mark options={solid, mycolor1}]
  table[row sep=crcr]{%
0.199999999999989	3.58727906461675\\
0.400000000000006	5.19217423889285\\
0.599999999999994	10.0291842235796\\
0.800000000000011	27.8098336267121\\
1	53.521371719588\\
2	132.576203580447\\
3	160.739331995789\\
};
\addlegendentry{Retransmit (Power Control)}

\addplot [color=mycolor2, dashdotted, line width=2.0pt, mark size=3.0pt, mark=x, mark options={solid, mycolor2}]
  table[row sep=crcr]{%
0.199999999999989	3.43883433518195\\
0.400000000000006	4.90003688309392\\
0.599999999999994	11.906596962759\\
0.800000000000011	42.8224772484936\\
1	68.4158548056414\\
2	121.66526660316\\
3	136.577974890298\\
};
\addlegendentry{Loopback (Power Control)}

\addplot [color=mycolor3, dotted, line width=2.0pt, mark=o, mark options={solid, mycolor3}]
  table[row sep=crcr]{%
0.200000000000003	71.1393579855692\\
0.400000000000006	81.8089019061437\\
0.599999999999994	90.3471064798454\\
0.799999999999997	102.732583179609\\
1	120.759908696176\\
2	120.878903649778\\
3	121.973471485429\\
};
\addlegendentry{Defer (Power Control)}

\addplot [color=black, line width=2.0pt, mark=diamond, mark options={solid, black}]
  table[row sep=crcr]{%
0.200000000000003	3.39269599191947\\
0.400000000000006	4.77402250739092\\
0.599999999999994	11.3730919452373\\
0.799999999999997	27.621384068711\\
1	56.0575907193337\\
2	120.789425331952\\
3	126.914137802096\\
};
\addlegendentry{IDA}

\end{axis}
\end{tikzpicture}%
\caption{IDA Mean Sojourn Times vs Arrival Rate}
\label{fig:IDA_pc}
\end{subfigure}

\begin{subfigure}{0.45\textwidth}
\vspace{20pt}
\centering
\input{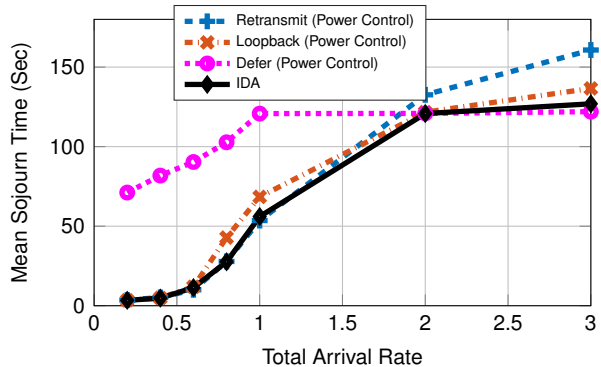}
\caption{IDA Convergence of Mean Sojourn Time. }
\label{fig:ida_conv}
\end{subfigure}

\begin{subfigure}{0.45\textwidth}
\vspace{20pt}
\centering
\input{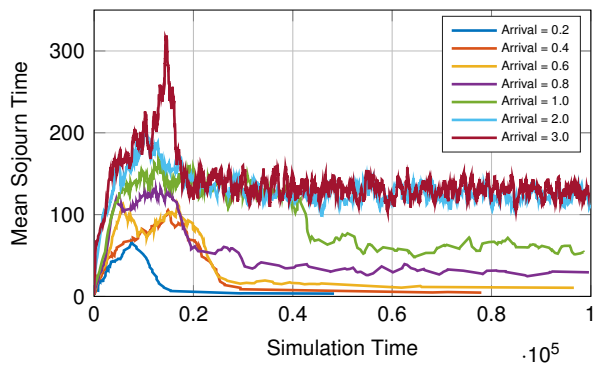}
\caption{IDA Average Power Convergence}
\label{fig:ida_pow_conv}
\end{subfigure}

\begin{subfigure}{0.45\textwidth}
\vspace{20pt}
\centering
\input{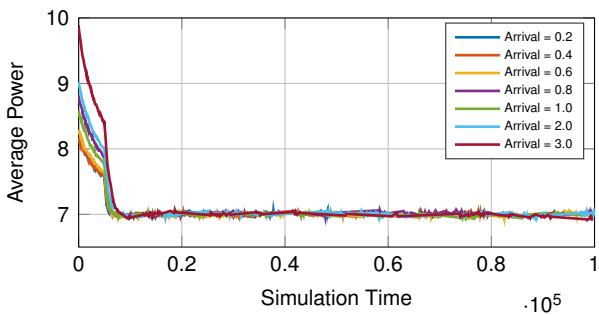}
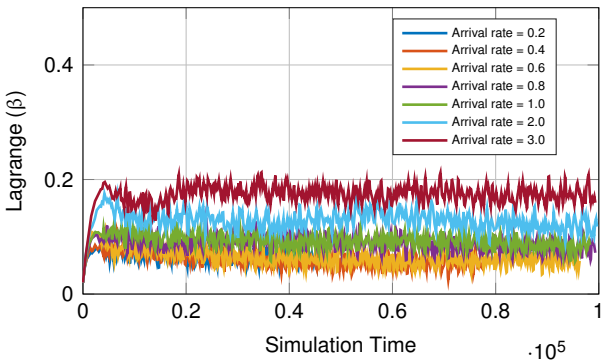
\caption{IDA Lagrange Convergence}
\label{fig:ida_beta_conv}
\end{subfigure}

\vspace{20pt}
\caption{IDA Performance in parametrized multicast system with $L=10$, $\overline{P}=7$, Zipf Popularity (Zipf exponent =1), Rayleigh fading with mean, 0.1 and 1.0 for bad and good users respectively.}

\end{figure}

\begin{figure}
\centering
\noindent\begin{subfigure}{0.45\textwidth}
\centering
% This file was created by matlab2tikz.
%
%The latest updates can be retrieved from
%  http://www.mathworks.com/matlabcentral/fileexchange/22022-matlab2tikz-matlab2tikz
%where you can also make suggestions and rate matlab2tikz.
%
\definecolor{mycolor1}{rgb}{0.00000,0.44700,0.74100}%
\definecolor{mycolor2}{rgb}{0.85000,0.32500,0.09800}%
\definecolor{mycolor3}{rgb}{0.92900,0.69400,0.12500}%
\begin{tikzpicture}

\begin{axis}[%
width=2.7in,
height=1.2in,
at={(0in,0in)},
scale only axis,
xmin=0,
xmax=100000,
xlabel={Simulation Time},
ymin=0,
ymax=1,
ylabel={Porbability Vector},
axis background/.style={fill=white},
xmajorgrids,
ymajorgrids,
legend style={nodes={scale=0.7, transform shape}, at={(0.3,0.4)}, anchor=south west, legend cell align=left, align=left, draw=white!15!black}
]
\addplot [color=mycolor1, dashdotted, line width=1.5pt]
  table[row sep=crcr]{%
0	0.400001574453199\\
1500	0.41996240496519\\
3600	0.373042317078216\\
4200	0.373257479513995\\
4900	0.394666150357807\\
6000	0.455262860006769\\
7600	0.483705988517613\\
8600	0.531199638237013\\
9600	0.576777244539699\\
10700	0.623733975706273\\
11600	0.679425021837233\\
12200	0.729492857470177\\
12600	0.78314714791486\\
13300	0.898501963369199\\
13900	1\\
29400	0.991612234400236\\
30600	0.987437851727009\\
31700	0.971174630554742\\
34200	0.915150595130399\\
40400	0.892129840751295\\
42200	0.876497480508988\\
43800	0.866902080524596\\
45400	0.851775694944081\\
46900	0.862389486923348\\
47800	0.850618484124425\\
48600	0.865396192195476\\
49600	0.872717415491934\\
55600	0.845010110730072\\
56600	0.852250382915372\\
58100	0.842367817924242\\
60700	0.860957318189321\\
62100	0.854677158669801\\
63600	0.860933133866638\\
64300	0.841509613077505\\
64900	0.830683180945925\\
65600	0.842366859636968\\
66800	0.858931142953224\\
68000	0.842120224915561\\
69000	0.83393132952915\\
71800	0.845872043602867\\
73200	0.841178571645287\\
75800	0.855983343499247\\
80900	0.841966818115907\\
83100	0.845564916889998\\
84800	0.840842644887744\\
86000	0.828609005431645\\
87400	0.825477359176148\\
89000	0.829399406924495\\
90300	0.830972818090231\\
92200	0.839809300072375\\
93900	0.814077631395776\\
96700	0.835839945575572\\
97600	0.8301170685736\\
};
\addlegendentry{Retransmit}

\addplot [color=mycolor2, dotted, line width=1.5pt]
  table[row sep=crcr]{%
0	0.415978700795677\\
1400	0.440002294708393\\
1800	0.416990461584646\\
2500	0.356559920372092\\
5700	0\\
29400	0.00838776559976395\\
30600	0.0125621482729912\\
31700	0.0288253694452578\\
34200	0.0848494048696011\\
40400	0.107870159248705\\
42200	0.123502519491012\\
43800	0.133097919475404\\
45400	0.148224305055919\\
46900	0.137610513076652\\
47800	0.149381515875575\\
48600	0.134603807804524\\
49600	0.127282584508066\\
55600	0.154989889269928\\
56600	0.147749617084628\\
58100	0.157632182075758\\
60700	0.139042681810679\\
62100	0.145322841330199\\
63600	0.139066866133362\\
64300	0.158490386922495\\
64900	0.169316819054075\\
65600	0.157633140363032\\
66800	0.141068857046776\\
68000	0.157879775084439\\
69000	0.16606867047085\\
71800	0.154127956397133\\
73200	0.158821428354713\\
75800	0.144016656500753\\
80900	0.158033181884093\\
83100	0.154435083110002\\
84800	0.159157355112256\\
86000	0.171390994568355\\
87400	0.174522640823852\\
89000	0.170600593075505\\
90300	0.169027181909769\\
92200	0.160190699927625\\
93900	0.185922368604224\\
96700	0.164160054424428\\
97600	0.1698829314264\\
};
\addlegendentry{Loopback}

\addplot [color=mycolor3, line width=1.5pt]
  table[row sep=crcr]{%
0	0.184019724765676\\
1300	0.134241014748113\\
1600	0.152298737957608\\
2100	0.205236712077749\\
4500	0.489842631766805\\
5000	0.526374298497103\\
5600	0.550183264902444\\
6000	0.544737139993231\\
7600	0.516294011482387\\
8600	0.468800361762987\\
9600	0.423222755460301\\
10700	0.376266024293727\\
11600	0.320574978162767\\
12200	0.270507142529823\\
12600	0.21685285208514\\
13300	0.101498036630801\\
13900	0\\
97600	0\\
};
\addlegendentry{Defer}

\end{axis}
\end{tikzpicture}%
\caption{Arrival rate=0.8}
\label{fig:ida_prob_0p8}
\end{subfigure}

\begin{subfigure}{0.45\textwidth}
\vspace{20pt}
\centering
% This file was created by matlab2tikz.
%
%The latest updates can be retrieved from
%  http://www.mathworks.com/matlabcentral/fileexchange/22022-matlab2tikz-matlab2tikz
%where you can also make suggestions and rate matlab2tikz.
%
\definecolor{mycolor1}{rgb}{0.00000,0.44700,0.74100}%
\definecolor{mycolor2}{rgb}{0.85000,0.32500,0.09800}%
\definecolor{mycolor3}{rgb}{0.92900,0.69400,0.12500}%
\begin{tikzpicture}

\begin{axis}[%
width=2.7in,
height=1.2in,
at={(0in,0in)},
scale only axis,
xmin=0,
xmax=100000,
xlabel={Simulation Time},
ymin=0,
ymax=1,
ylabel={Porbability Vector},
axis background/.style={fill=white},
xmajorgrids,
ymajorgrids,
legend style={nodes={scale=0.7, transform shape}, at={(0.45,0.7)}, anchor=south west, legend cell align=left, align=left, draw=white!15!black}
]
\addplot [color=mycolor1, dashdotted, line width=1.5pt]
  table[row sep=crcr]{%
0	0.196066935663112\\
1900	0.126919966744026\\
2500	0.133803585355054\\
2900	0.156771418216522\\
3400	0.208517896724516\\
4100	0.30223076956463\\
6300	0.614395397627959\\
6900	0.66483283128764\\
7400	0.689726134558441\\
8100	0.704625033773482\\
8800	0.701642024010653\\
9900	0.671764365761192\\
10800	0.640031772563816\\
11500	0.561096256584278\\
12800	0.42003647920501\\
14000	0.305040893159457\\
15400	0.168232611467829\\
16400	0.0511729039862985\\
16800	0\\
97900	0\\
};
\addlegendentry{Retransmit}

\addplot [color=mycolor2, dotted, line width=1.5pt]
  table[row sep=crcr]{%
0	0.33815827453509\\
1700	0.330857236054726\\
2500	0.305003701956593\\
3100	0.270284319776692\\
4400	0.171227468192228\\
5700	0.0748881140607409\\
6400	0.0390284662862541\\
7000	0.0231041832012124\\
7500	0.0263189737743232\\
8100	0.0490533606207464\\
8600	0.0841887458227575\\
9200	0.145183889340842\\
10200	0.27097730548121\\
11400	0.427401065637241\\
12800	0.57996352079499\\
14000	0.694959106840543\\
15400	0.831767388532171\\
16400	0.948827096013702\\
16800	1\\
29700	0.991818420443451\\
30700	0.971095750559471\\
36600	0.790463762023137\\
39700	0.668080026545795\\
42600	0.554764280954259\\
49000	0.352443862939253\\
50800	0.321825191727839\\
55400	0.289409817793057\\
57300	0.300340861504083\\
59000	0.32678710516484\\
64700	0.461987276969012\\
69800	0.602629946253728\\
72300	0.647892068547662\\
74500	0.662027554295491\\
78300	0.682860370739945\\
80300	0.702585488528712\\
82400	0.721012262787553\\
86000	0.715189529015333\\
88200	0.708865735898144\\
90700	0.709398499384406\\
92500	0.712846186739625\\
94600	0.714455679175444\\
97900	0.721920933487127\\
};
\addlegendentry{Loopback}

\addplot [color=mycolor3, line width=1.5pt]
  table[row sep=crcr]{%
0	0.465774789801799\\
2600	0.562150191923138\\
3100	0.554680149260093\\
3800	0.521438531606691\\
4700	0.459045481024077\\
6500	0.331338062242139\\
8600	0.211366173287388\\
9600	0.126512714879937\\
11000	0\\
29700	0.00818157955654897\\
30700	0.0289042494405294\\
36600	0.209536237976863\\
39700	0.331919973454205\\
42600	0.445235719045741\\
49000	0.647556137060747\\
50800	0.678174808272161\\
55400	0.710590182206943\\
57300	0.699659138495917\\
59000	0.67321289483516\\
64700	0.538012723030988\\
69800	0.397370053746272\\
72300	0.352107931452338\\
74500	0.337972445704509\\
78300	0.317139629260055\\
80300	0.297414511471288\\
82400	0.278987737212447\\
86000	0.284810470984667\\
88200	0.291134264101856\\
90700	0.290601500615594\\
92500	0.287153813260375\\
94600	0.285544320824556\\
97900	0.278079066512873\\
};
\addlegendentry{Defer}

\end{axis}
\end{tikzpicture}%
\caption{Arrival rate=2.0}
\label{fig:ida_prob_2p0}
\end{subfigure}

\begin{subfigure}{0.45\textwidth}
\vspace{20pt}
\centering
% This file was created by matlab2tikz.
%
%The latest updates can be retrieved from
%  http://www.mathworks.com/matlabcentral/fileexchange/22022-matlab2tikz-matlab2tikz
%where you can also make suggestions and rate matlab2tikz.
%
\definecolor{mycolor1}{rgb}{0.00000,0.44700,0.74100}%
\definecolor{mycolor2}{rgb}{0.85000,0.32500,0.09800}%
\definecolor{mycolor3}{rgb}{0.92900,0.69400,0.12500}%
\begin{tikzpicture}

\begin{axis}[%
width=2.7in,
height=1.2in,
at={(0in,0in)},
scale only axis,
xmin=0,
xmax=100000,
xlabel={Simulation Time},
ymin=0,
ymax=1,
ylabel={Porbability Vector},
axis background/.style={fill=white},
xmajorgrids,
ymajorgrids,
legend style={nodes={scale=0.7, transform shape}, at={(0.4,0.5)}, anchor=south west, legend cell align=left, align=left, draw=white!15!black}
]
\addplot [color=mycolor1, dashdotted, line width=1.5pt]
  table[row sep=crcr]{%
0	0.324992223846493\\
2200	0.310503890636028\\
3400	0.294588359596673\\
4100	0.308559837678331\\
4600	0.336003879187047\\
5500	0.411728570601554\\
6500	0.498651403511758\\
8600	0.542018682783237\\
9100	0.574282933579525\\
9700	0.635055933642434\\
10400	0.729379587020958\\
11700	0.915487559439498\\
11900	0.939839661077713\\
12200	0.919235052046133\\
12800	0.855307570513105\\
13400	0.773672375929891\\
14200	0.643072527120239\\
15200	0.458808259441867\\
16900	0.144436727263383\\
17700	0.00602945554419421\\
17900	0\\
97600	0\\
};
\addlegendentry{Retransmit}

\addplot [color=mycolor2, dotted, line width=1.5pt]
  table[row sep=crcr]{%
0	0.28793215440237\\
2000	0.266396593156969\\
2500	0.278997559507843\\
3400	0.327549568246468\\
4100	0.385030611156253\\
4700	0.431429195727105\\
5600	0.476380518521182\\
6500	0.501348596488242\\
8600	0.457981317216763\\
9100	0.425717066420475\\
9700	0.364944066357566\\
10300	0.283701338194078\\
11000	0.169596536856261\\
12000	0\\
23300	0.00894706210237928\\
25400	0.030764884679229\\
27100	0.0425814340414945\\
29400	0.0440589409845416\\
30400	0.0240708324708976\\
33100	0\\
97600	0\\
};
\addlegendentry{Loopback}

\addplot [color=mycolor3, line width=1.5pt]
  table[row sep=crcr]{%
0	0.387075621751137\\
2300	0.419752216199413\\
2900	0.40498653704708\\
3400	0.377862072142307\\
3900	0.330863679206232\\
4600	0.238979130779626\\
6000	0.0524606619292172\\
6500	0\\
10900	0.0089029109076364\\
11500	0.0305172046646476\\
12100	0.0721732933452586\\
12600	0.121695849753451\\
13200	0.197022530264803\\
13900	0.305853575540823\\
14700	0.446978331412538\\
17500	0.961229444728815\\
17700	0.993970544455806\\
17900	1\\
23300	0.991052937897621\\
25400	0.969235115320771\\
27100	0.957418565958505\\
29400	0.955941059015458\\
30400	0.975929167529102\\
33100	1\\
97600	1\\
};
\addlegendentry{Defer}

\end{axis}
\end{tikzpicture}%
\caption{Arrival rate=3.0}
\label{fig:ida_prob_3p0}
\end{subfigure}

\vspace{20pt}
\caption{IDA convergence of queueing strategies for different arrival rates. $L=10$, $\overline{P}=7$, Zipf Popularity (Zipf exponent =1), Rayleigh fading with mean, 0.1 and 1.0 for bad and good users respectively.}
\end{figure}
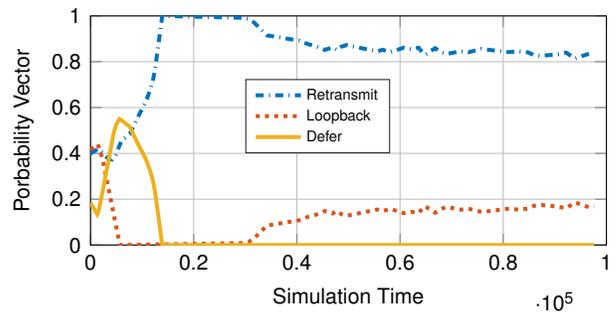
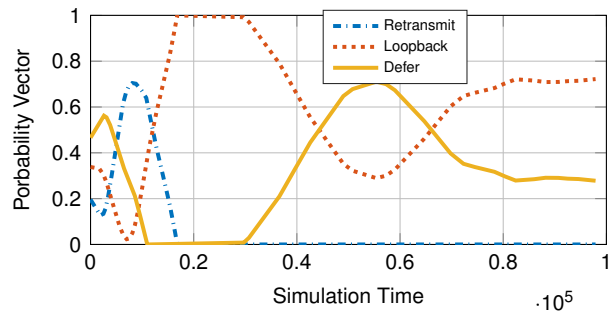
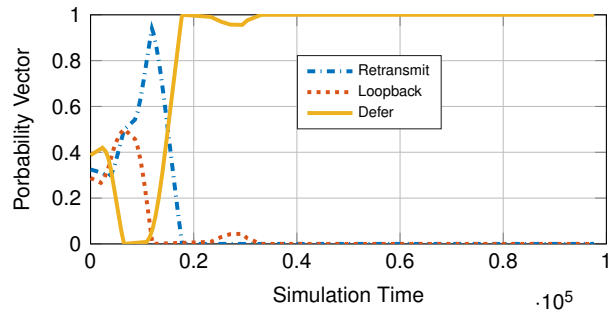

We have already seen the performance of power control for 1LB (Loopback case) for large user system. In this section we compare the performances of AC-DQN for different queueing strategies versus the IDA performance for the moderate user case (Section \ref{sec:system_10}). We use Zipf popularity and Rayleigh fading for system simulation. First, in Figure \ref{fig:IDA_pc}, we make an observation that AC-DQN drastically improves the mean delay performance for all the strategies as compared to the constant power policy in Figure \ref{fig:dsgd_cp}. We see that our IDA algorithm is able to choose better strategy than the baselines in terms of mean sojourn time. The convergence of mean sojourn time for rates $0.2$ to $3.0$ is shown in Figure \ref{fig:ida_conv}. The more important capability of this algorithm is that it \CE converges to a better mean sojourn time \CE while maintaining the average power constraint. Figure \ref{fig:ida_pow_conv} shows convergence of the average power to $\overline{P}=7$ for all the rates. This is achieved by simultaneously controlling the Lagrange variable as seen in Figure \ref{fig:ida_beta_conv}. A few interesting plots showing convergence of probabilities for rates $0.8,\ 2.0$ and $3.0$ are shown in Figures \ref{fig:ida_prob_0p8}, \ref{fig:ida_prob_2p0} and \ref{fig:ida_prob_3p0} respectively. 

		\CE We see, in \ref{fig:ida_prob_0p8}, for arrival rate $0.8$, that the probability converges to a mixed policy with $0.8$ probability assigned to retransmit and $0.2$ assigned to loopback, though we see in Figure \ref{fig:IDA_pc} that retransmit individually has the same optimal mean sojourn time as achieved by IDA\CE. This is attributed to more flexibility available with the algorithm. If the DeepRL in IDA finds an optimal power control policy along with this mixed policy that achieves the optimal mean delay, it may converge to that policy. In other words this flexibility gives additional optimal points for the algorithm, to choose from. Infact, we see that in Figure \ref{fig:ida_prob_0p8} that the algorithm first goes to retransmit and eventually converges to this mixed policy, while maintaining average power and optimal delay throughout. For rate $0.2$ both defer and loopback have same AC-DQN performance Figure \ref{fig:IDA_pc}. Hence, the solution oscillates between them Figure \ref{fig:ida_prob_2p0}, while maintaining the average power and optimal delay. For arrival rate $3.0$, Figure \ref{fig:ida_prob_3p0}, however the algorithm unambiguously chooses defer as the policy since it has \CE the lowest mean sojourn time among the baselines, Fig \ref{fig:IDA_pc}. The simulations show that IDA is able to achieve adaptive cross-layer optimization of queueing and power control simultaneously, for different system statistics (arrival rates)\CE.

%
%\begin{figure}[h!]
%\centering
%%\includegraphics[trim={3.5cm 9cm 3.5cm 9.5cm},clip,height=4cm,width=8cm]{JournalFigures/IDA_Prob_0p8.pdf}
%\input{IDA_Prob_0p8.tex}
%\caption{IDA Probability convergence with Time for arrival rate 0.8. $L=10$, $\overline{P}=7$, Zipf Popularity (Zipf exponent =1), Rayleigh fading with mean, 0.1 and 1.0 for bad and good users respectively.}
%\label{fig:ida_prob_0p8}
%\end{figure}
%
%\begin{figure}[h!]
%\centering
%%\includegraphics[trim={3.5cm 9cm 3.5cm 9.5cm},clip,height=4cm,width=8cm]{JournalFigures/IDA_Prob_2p0.pdf}
%\input{IDA_Prob_2p0.tex}
%\caption{IDA Probability convergence with Time for arrival rate 2.0. $L=10$, $\overline{P}=7$, Zipf Popularity (Zipf exponent =1), Rayleigh fading with mean, 0.1 and 1.0 for bad and good users respectively.}
%\label{fig:ida_prob_2p0}
%\end{figure}
%
%\begin{figure}[h!]
%\centering
%%\includegraphics[trim={3.5cm 9cm 3.5cm 9.5cm},clip,height=4cm,width=8cm]{JournalFigures/IDA_Prob_3p0.pdf}
%\input{IDA_Prob_3p0.tex}
%\caption{IDA Probability convergence with Time for arrival rate 3.0. $L=10$, $\overline{P}=7$, Zipf Popularity (Zipf exponent =1), Rayleigh fading with mean, 0.1 and 1.0 for bad and good users respectively.}
%\label{fig:ida_prob_3p0}
%\end{figure}

\subsection{Discussion:} We see from the simulations that the novel Deep Learning techniques such as DSGD and AC-DQN can achieve optimal performance while providing scalability with system size. We have demonstrated how DNNs can be used for noise reduction in gradient estimates in a stochastic gradient descent algorithm, as done in DSGD. Our two-timescale approach,  AC-DQN, extends DeepRL algorithms like DQN to systems with constrained control. Though we have demonstrated this on a system with a single constraint, it can be extended to systems with multiple constraints. In such systems each constraint is associated with a Lagrange multiplier. Each Lagrange multiplier adds an additional SGD step to the AC-DQN algorithm. For a stationary system, it is enough that the step-sizes satisfy multi-timescale criterion similar to (\ref{eq:two_timeline_learning}), see \cite{borkar}. However, if AC-DQN is used in systems with changing system statistics the step sizes shall be kept constant. The step sizes shall be fixed as per the tolerance requirement for a given constraint (e.g., in our system the tolerance could be $\overline{P}\pm \Delta P$, where, $\Delta P$ is the allowed deviation from the constraint $\overline{P}$). Lesser the tolerance, lesser the step-size. However, fixing the step-sizes too small may make the algorithm too slow to track the changes in system statistics. Hence, choosing the step sizes is a trade-off between the tolerance of the constraint and the required algorithmic agility to track the system changes. We have shown that this Multi-timescale approach in AC-DQN can very well be extended to systems with multiple objectives as demonstrated by our IDA algorithm. We have also demonstrated how IDA achieves the optimal queuing strategy among the baselines while obtaining the power control for such complex multicast systems. It is shown that Deep Neural Networks when appropriately used can provide scalable control for large wireless networks. Infact it can simultaneously achieve several cross-layer objectives even in large wireless networks for providing optimal QoS.   

\section{Conclusion}
\label{sec:conclusion}

\CE This paper considered a multicast downlink in single hop wireless network. Fading of different links to users causes significant reduction in the performance of the system. Appropriate  change in the queueing policies and power control can mitigate most of the losses.  \CE However, simultaneously obtaining adaptive queueing and power control for large systems is computationally very hard\CE. We first develop a novel DNN assisted stochastic gradient descent algorithm to achieve optimality of the system to provide lower mean sojourn time in a parametrized multicast system. Next we show that using Deep Reinforcement Learning, we can obtain optimal power control, online, even when the system statistics are unknown. We use a recently developed version  of Q learning, Deep Q Network to learn the Q-function of the system via function approximation. Furthermore, we modify the algorithm to satisfy our constraints and also to make the optimal policy track the time varying system statistics. Finally, we propose a novel deep multi-time scale algorithm which achieves the cross-layer optimization of queuing and power control, simultaneously. 

One interesting extension of this work would be developing an algorithm that could potentially provide better state-action dependent queueing strategy. Another future work could possibly include, the caches at the user nodes and learning the optimal caching policy along with the power control using DeepRL. Future works may also consider applying IDA to multiple-base-station scenarios for interference mitigation. \CE

\bibliographystyle{IEEEtran}
\bibliography{library}
\end{document}